%% file: ms2.tex
\documentclass[manuscript]{aastex}

\newcommand{\feh}{\mathrm{[Fe/H]}}
\newcommand{\teff}{T_\mathrm{eff}}
\newcommand{\logg}{\log g}
\newcommand{\afe}{A_\mathrm{Fe}}
\newcommand{\fei}{Fe\,\textsc{i}}
\newcommand{\feii}{Fe\,\textsc{ii}}
\newcommand{\kms}{km\,s$^{-1}$}
\newcommand{\ha}{H$\alpha$}

\slugcomment{Submitted to The Astrophysical Journal.}

\shorttitle{Gyrochronology with wide binaries}
\shortauthors{Chanam\'e \& Ram\'{i}rez}

\begin{document}


\title{Towards precise ages for single stars in the field.\\
  Gyrochronology constraints at several Gyr using wide binaries.\\
  I. Ages for initial sample}


\author{Julio Chanam\'e\altaffilmark{1}}
\affil{Department of Terrestrial Magnetism, Carnegie Institution of Washington\\
5241 Broad Branch Road N.W., Washington DC, 20015\\
and\\
Departamento de Astronom\'{i}a y Astrof\'{i}sica, Pontificia
Universidad Cat\'olica de Chile\\
Av. Vicu\~na Mackenna 4860, 782-0436 Macul, Santiago, Chile
}

\author{Iv\'an Ram\'{i}rez\altaffilmark{2}}
\affil{The Observatories of the Carnegie Institution for Science\\
813 Santa Barbara Street, Pasadena, CA 91101
}
%
%


\altaffiltext{1}{Hubble Fellow}
\altaffiltext{2}{Sagan Fellow}


\begin{abstract}

  We present a program designed to obtain age-rotation measurements of
  solar-type dwarfs to be used in the calibration of gyrochronology
  relations at ages of several Gyr. This is a region of parameter
  space crucial for the large-scale study of the Milky Way, and where
  the only constraint available today is that provided by the Sun. Our
  program takes advantage of a set of wide binaries selected so that
  one component is an evolved star and the other is a main-sequence
  star of FGK type.  In this way, we obtain the age of the system from
  the evolved star, while the rotational properties of the main
  sequence component provides the information relevant for
  gyrochronology regarding the spin-down of solar-type stars. By
  mining currently available catalogs of wide binaries, we assemble a
  sample of 37 pairs well positioned for our purposes: 19 with turnoff
  or subgiant primaries, and 18 with white dwarf components.  Using
  high-resolution optical spectroscopy, we measure precise stellar
  parameters for a subset of 15 of the pairs with turnoff/subgiant
  components, and use these to derive isochronal ages for the
  corresponding systems. Ages for 16 of the 18 pairs with white dwarf
  components are taken from the literature. The ages of this initial
  sample of 31 wide binaries range from 1 to 9 Gyr, with precisions
  better than $\sim 20$\% for almost half of these systems. When
  combined with measurements of the rotation period of their main
  sequence components, these wide binary systems would potentially
  provide a similar number of points useful for the calibration of
  gyrochronology relations at very old ages.


\end{abstract}


\section{Introduction}\label{intro}

Understanding the detailed formation and evolution of systems such as
star clusters and galaxies requires knowledge of the ages of their
constituent stars with sufficient precision. The derived star
formation histories of the populations of stars in all types of
galaxies provide crucial constraints on models of the formation of
structure on cosmological scales \citep{spr03,nag05,con09}. In the
context of our own Milky Way, for example, the various theoretical
models for the formation and evolution of the disk of the Galaxy make
different predictions for the age distributions of thin- and
thick-disk stars \citep{aba03,bro07,sch09}. Even though large amounts
of information on the chemistry and kinematics of these objects are
and will continue to become available, without a proper knowledge of
the ages of thin- and thick-disk members it will not be possible to
determine which are the more realistic scenarios. Similarly,
age-dating of the numerous stellar streams being discovered by modern
surveys would help trace back their origin and therefore contribute to
a better understanding of the recent accretion history of the Galaxy.
Exoplanet research could also benefit from precise determinations of
stellar ages since the ages of host stars can be used as constrains
for dynamical models of planetary systems and migration. Moreover, the
ages of host stars would be crucial for investigations of biological
evolution in potentially habitable planets \citep{sod10}.

Unfortunately, the determination of stellar ages is possible only in a
few cases, typically for sets of stars that can be safely assumed to
be more or less coeval (e.g., clusters, associations, etc.) and for
very nearby stars for which a wealth of accurate information such as
distances, luminosities, and detailed chemical compositions are also
available. For plain and simple isolated stars in the Galactic field
just outside the very immediate solar neighborhood, reliable ages are
simply not possible yet. This lack of knowledge of the age
distribution of stars in the Milky Way thus constitutes a severe
limitation under the perspective of cosmology and galaxy formation, as
it prevents an adequate investigation of the details of the formation
history of our Galaxy. Upcoming experiments such as Gaia and LSST aim
to change this situation by measuring the properties of many millions
of stars with very high accuracy but, as of today, stellar ages will
still need to be derived from isochrone fitting. While historically
this has been the most successful method for deriving stellar ages in
many branches of astronomy, it suffers from limitations that become
particularly restrictive when dealing with the needs of a more
ambitious mapping of the star formation histories of the components of
the Milky Way. In this context, some of the limitations of isochronal
ages include their high sensitivity to errors in distance
\citep{sod05} and stellar parameters, systematic biases arising from
the sampling of isochrone points
\citep{pont04,nordstrom04,jorgensen05,dasilva06}, the impact of poorly
understood stellar evolutionary processes such as microscopic
diffusion, convective overshoot, gravitational settling, and others
\citep{vandenberg02,demarque04,michaud10}, and the unavoidable fact
that the method is most sensitive to evolved, thus relatively massive,
stars, and not so much to stars on the main sequence (MS). Thus the
importance of seriously exploring any alternative to the use of
stellar models for deriving ages of field stars.

One of such alternatives is that offered by gyrochronology
\citep{sku72,bar03,bar07}. The method takes advantage of the fact that
MS stars of FGK-types are known to lose angular momentum in a
predictable way, as seen by the measured surface rotation rates of
stars in a sequence of clusters of different ages
\citep{stau87,rad95,irwin06,messina08,meibom09,collier09,hartman09,
hartman10,delorme11,meibom11a}. Thus, if it were possible to map and
quantitatively calibrate with precision such spin down from very young
to very old ages, it should in principle be possible to apply those
gyrochronology relations to single stars in the field and, by
measuring their surface rotation rates and colors (a proxy for their
masses), readily obtain their ages.

While there exists a growing body of observational evidence in support
of the possibility of gyrochronology, a solid theoretical
understanding still needs to be fully developed. The physics at the
heart of the method involves the self-regulated interaction between
stellar phenomena that are not easily modeled, such as differential
internal rotation, magnetic fields interior to stars, and magnetized
stellar winds that are able to carry angular momentum away. For a
discussion of these processes and their interaction in the context of
a theoretical framework for gyrochronology, see \citet{bar03,bar10}.
In the present paper, we approach the subject from the perspective of
the empirical calibration of gyrochronology relations for use as a
tool for the age dating of field stars.

Until very recently, reliably calibrated gyrochronology relations
existed only for stars younger than about 0.5 Gyr, for the simple
reason that such age corresponded to the oldest clusters for which
rotation periods had been measured for a large enough number of member
stars.  The only constraint beyond this point corresponded to the Sun,
thus preventing a good calibration of gyrochronology relations for
stars a few Gyr old.  Although there was never a lack of older
clusters to extend these relations to correspondingly older ages, the
limitation was of a technical nature as older stars become
progressively less active, thus complicating the task of measuring the
already small photometric modulation due to starspots coming into and
out of the line-of-sight as the stars rotate.

The above situation started to improve recently when \citet{meibom11b}
published rotation periods for stars in the 1 Gyr old cluster NGC
6811.  Theirs is a comprehensive program that takes advantage of the
existence of three open clusters within the field of view of NASA's
{\it Kepler} mission, and thus it is expected that their work for NGC
6811 will be repeated for the cases of NGC 6819 and NGC 6791, which
would provide calibrations of gyrochronology relations at 2.5 and,
possibly, 9 Gyr of age, respectively.

In this paper we introduce an ongoing program devised to place
constraints for gyrochronology relations in the regime of ages where
the large majority of field stars fall, i.e., from one to several Gyr.
We achieve this by studying samples of wide stellar binaries chosen so
that one of their components is an evolved star (either turnoff,
subgiant, or white dwarf), while their secondaries are regular MS
stars of FGK types.  Therefore, with the evolved primary potentially
well suited to provide a reliable age for the system\footnote{The
  assumption of a common age for the components of a wide binary
  should be a safe one.  In the context of the recently proposed
  formation mechanism of \citet{kou10}, where wide binaries are formed
  during the dissolution phase of young star clusters when pairs of
  initially unassociated stars get close in phase space and become
  bound, the maximum difference possible between the ages of the two
  components cannot be larger than the age of the dissolving young
  cluster (on the order of $10^6-10^8$ yr), thus small,
  when not negligible, in comparison with the present Gyr-scale ages
  of these systems.}, a measurement of the rotation period of the MS
secondary automatically provides a point useful for the calibration of
gyrochronology relations at the corresponding age.

The availability of large numbers of wide binaries with the above
characteristics, both from existing as well as upcoming all-sky
surveys and catalogs, suggests that these objects could in principle
provide quite a large and varied pool of gyrochronology constraints.
Indeed, being a representative sample of the mix of stellar
populations that surround the solar neighborhood, wide binaries span a
large range of stellar properties, including age, mass, and
metallicity. For these reasons, a gyrochronology program based on
these objects has the potential of populating the phase space of
relevant variables (i.e., age, mass, and rotation period) more densely
and continuously than it may be possible with only star cluster
observations.

The first step in our program, therefore, involves the determination
of stellar ages for a sample of wide binary components suitable of
being dated. Today, precise stellar ages for individual stars can be
readily computed for stars that either (a) began relatively recently
to evolve away from the MS, such as turnoff stars and subgiants, or
(b) did this a long time ago and are now in the white dwarf (WD) phase
of their evolution. For recently evolved stars, reliable ages are
routinely obtained via the use of adequately-chosen isochrones,
provided stellar parameters such as effective temperature ($T_{\rm
  eff}$), metallicity ([Fe/H]), surface gravity ($\log g$), and
distance are known with precision. In the case of a WD, if it is of a
DA type (i.e., with a hydrogen-dominated atmosphere), its cooling time
($t_{\rm cool}$) and mass can be derived from its $T_{\rm eff}$ and
$\log g$ (measured from its spectrum) and the use of appropriate
cooling models. Then, using empirically-calibrated initial-to-final
mass relations one can derive the mass and lifetime of the progenitor
star, which, when added to $t_{\rm cool}$ provides the total age of
the WD \citep{zhao11,garces11}. Thus, wide binary systems where one of
the components is of either of these types is suitable to be
age-dated, and this is what we plan to take advantage of in order to
assign ages to their FGK components, which could then serve as
gyrochronology constraints.

Once their ages have been measured, the remaining step for obtaining
gyrochronology constraints requires the measurement of the rotation
periods of the MS components of these pairs. Rotation periods of
individual stars are typically obtained by monitoring the small
modulation of the star's brightness as spots come in and out of the
line-of-sight as the star rotates\footnote{Rotation periods obtained
in this way are thus independent of the inclination of the rotation
axis, i.e., there is no $\sin i$ ambiguity as in the case of the
measurement of the rotation velocity from the broadening of the star's
spectral lines.  Transforming from a rotation velocity to a rotation
period would suffer, additionally, from the uncertainty on the actual
size of the star.} (e.g., \citealt{hartman11}).  The monitoring can
also be done spectroscopically, by following the intensity variation
of the emission at the cores of the Ca II H and K lines as the star
rotates (e.g., \citealt{vaughan81,baliunas83,cincunegui07,hall07}).
We have not been able to find published rotation periods for any of
the MS components of the pairs presented in this paper.  These targets
are bright stars ($5 < V < 10$ mag), and typical stellar rotation
periods at such old ages are of the order of a month and longer.  Two
space missions, the MOST and CoRoT satellites, are very well posed for
this kind of work, routinely achieving precisions of a few
millimagnitudes and smaller on stars similar to those of our program
(e.g., \citealt{miller08,siwak10,strassmeier10,corot11}).  From the
ground, the only possible way to achieve this is using
almost-dedicated telescopes with small apertures, and for this purpose
we started to take advantage of instrumentation of that kind at the
Observatorio Docente of Universidad Cat\'olica in the outskirts of
Santiago, Chile.  This part of the program is currently in progress
for an initial sample of the systems presented here, and will be
reported upon completion in forthcoming papers.

In this paper we describe our selection of an initial sample of wide
binaries suitable for the purposes of this program and report on the
ages of a number of these systems, obtained either from our own
methods or directly from the literature. Section \ref{targets}
describes the selection of targets. In \S\,3 we describe our procedure
to determine accurate stellar parameters for a subset of target stars
to be used in the process of age determination, which is reported in
\S\,4. In \S\,5 we summarize our program and initial results.

\section{Sample Selection}\label{targets}

There exists a number of published wide binary catalogs that can be
used to select systems suitable for our gyrochronology program.  These
catalogs have been assembled from a variety of parent surveys and
following a variety of selection criteria regarding angular separation
limits, proper-motion cuts, stellar type, and even distances.  Since
the crucial assumption lying at the core of this program is that a
number of properties that can be well measured for one of the
components of a wide binary can be safely assigned to the other
component\footnote{For close binaries, of course, this statement might
not always be true, as their components may have interacted with each
other and thus affected their normal evolution.}, we need to maximize
the chances of selecting genuine, truly bound pairs of stars, avoiding
any possible contamination by unassociated pairs as much as possible.
Therefore, when searching for useful wide binaries for our program the
best sources will be those that provide the largest amount of
information (position, photometry, kinematics) that can be used to
assess the evidence for binarity on any given pair of stars.
Moreover, given that useful gyrochronology constraints require the
determination of reliable ages, the best targets for our program will
be those for which stellar parameters can be measured with high
precision.

The publicly available wide binary catalogs that best satisfy the
above requirements are those of Chanam\'e \& Gould (2004; hereafter
CG04), Gould \& Chanam\'e (2004; hereafter GC04) and L\'epine \&
Bongiorno (2007; hereafter LB07).  Although the literature offers a
few catalogs with larger numbers of binaries than these three, most
nevertheless suffer from shortcomings that would negatively impact a
sample aimed for gyrochronology: large contamination by the chance
alignment of unassociated pairs (e.g., the catalog of
\citealt{sesar08}, based on SDSS data), faint stars comprising the
large majority of objects in the catalog (e.g., the SLoWPoKES catalog
of \citealt{dhital10}, also based on SDSS data and which, moreover,
was aimed by construction at very low-mass, late-K and M stars), and
even a significant fraction of systems in the catalog expected to be
not bound anymore \citep{shaya11}.

In contrast, the CG04, GC04, and LB07 catalogs only contain pairs with
high probability of being genuinely bound systems, satisfying
stringent requirements not only of kinematical nature (i.e., both
stars displaying a common proper-motion), but also having luminosities
and colors consistent with the two stars sharing the same age and
chemical composition.  Moreover, unlike those based on SDSS data, the
CG04, GC04, and LB07 catalogs were assembled from surveys of high
proper-motion stars, where the probability of two stars being close in
phase space and moving at such high velocities but not being
associated is already very small.  Finally, having been selected from
a high proper-motion survey, the stars in these three catalogs are
among the brightest on the sky, and hence the best suited to provide
precise stellar parameters through high-resolution spectroscopic work.

Before proceeding to the selection of targets, we briefly discuss the
impact on our program of undetected higher-order multiplicity among
wide binaries \citep{correia06,tok10}. If the evolved star of a pair
in our sample has an unresolved, undetected close companion, in
principle the stellar parameters measured for the program star may be
affected by light from the undetected neighbor, and thus the age
derived for the system. However, as long as such hypothetical
companion is significantly fainter than the primary star, for high
signal-to-noise observations such as those we deal with in
\S\,\ref{ivan} the impact of an unresolved companion would be minimal.
Moreover, since we are working in the visible part of the spectrum,
the effect is even more negligible. The impact of a cool, close
companion on the observed spectrum of the evolved star would be more
important in the infrared and at longer wavelengths. On the other
hand, if the star with an unresolved companion is the FGK-type MS
member of the pair, then there is a chance, depending on how close the
unseen companion is, that the rotational properties of the program
star have been affected by interactions with such companion, and thus
its surface rotation rate may not reflect the processes behind
gyrochronology. This is a worry in common with gyrochronology programs
based on star clusters, and indeed many of them include a campaign of
spectroscopic monitoring in order to identify close binaries (e.g.,
\citealt{meibom09,meibom11a}). In conclusion, pairs with MS components
having signs of a close companion should be avoided and are thus
excluded from our program.

\subsection{Wide binaries with turnoff and subgiant components}

Given that isochronal ages are very sensitive to uncertainties in
distance, we restrict our selection to binary systems for which at
least one of the components is an {\it Hipparcos} star. This is the
case of all entries in the GC04 and LB07 catalogs, and all pairs in
CG04 with this characteristic are by construction already listed in
GC04. Therefore, in order to select wide binaries with recently
evolved components, we focus on the GC04 and LB07 catalogs and
restrict our search to pairs with better than $3\sigma$ parallaxes
($\pi/\sigma_{\pi} \geq 3$).

As a first step, for all entries in the GC04 and LB07 catalogs, we
make the distinction between pairs in which both components are {\it
  Hipparcos} stars with independent parallax measurements, and wide
binaries in which only one of the components is in {\it Hipparcos}.
This is simply because having independent trigonometric distances to
both stars in any given pair provides an important extra criterion for
assessing the true binary nature of the stellar pair, thus placing
these cases in a different category.

\begin{figure}
\epsscale{1.00}
\plotone{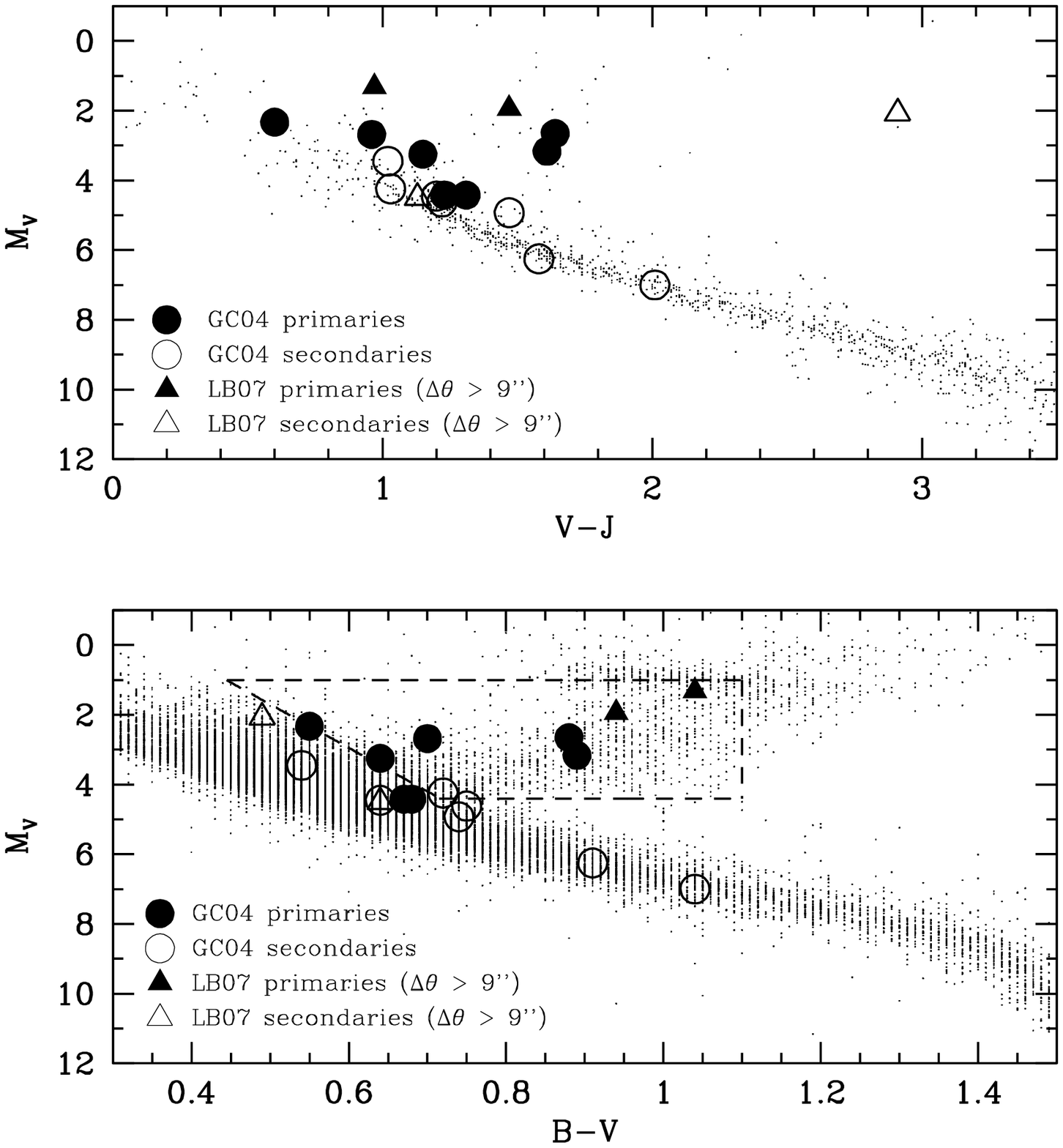}
\caption{Selection of wide binary pairs with turnoff/subgiant
  components, both members having independent {\it Hipparcos}
  parallaxes (primaries as filled symbols, secondaries as open
  symbols).  The 9 pairs in this category are shown here on two
  different color-magnitude diagrams and listed in Table
  \ref{tab:hipp_doubles}.  The small background dots in the $B-V$ CMD
  are Hipparcos single stars within 100 pc of the Sun and with better
  than $5−\sigma$ parallaxes, and in the $V-J$ CMD they are single
  Hipparcos stars in the LSPM catalog and within 33 pc of the Sun
  (L\'epine 2005). Primaries within the dashed area in the $B-V$ CMD
  and with main-sequence secondaries bluer than $B-V = 1.40$ are
  selected for further inspection. }.
\label{fig:sel1}
\end{figure}

\begin{figure}
\epsscale{1.00}
\plotone{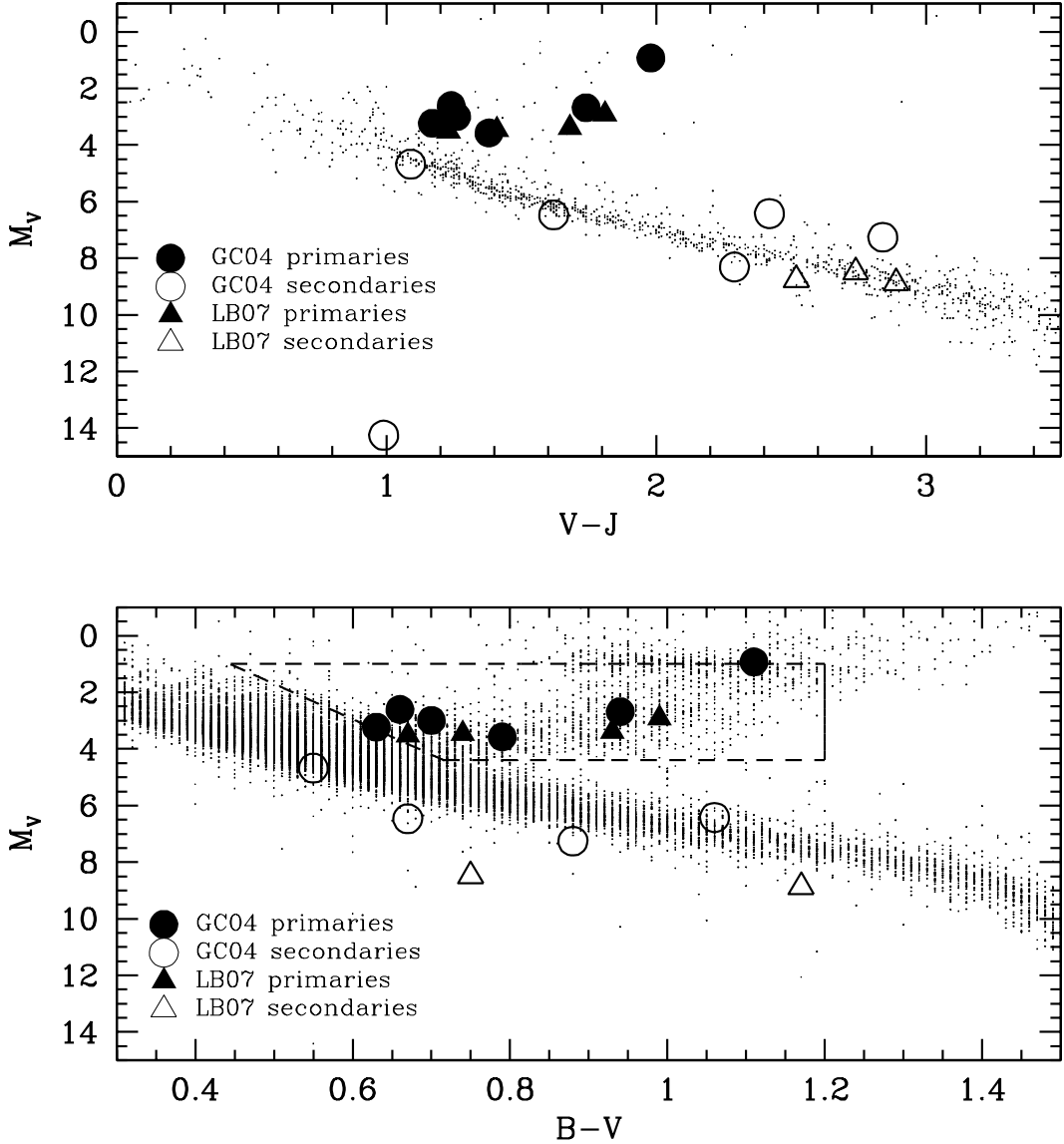}
\caption{Same as Figure \ref{fig:sel1}, but for pairs with only one
  {\it Hipparcos} component.  The 10 pairs shown here are listed in
  Table \ref{tab:hipp_nonhipp}.  Primaries within the dashed area in
  the $B-V$ color-magnitude diagram and with main-sequence secondaries
  bluer than $V-J \sim 3.0$, approximately the boundary between K and
  M dwarfs, are selected for further inspection.  The magnitudes of a
  few pairs have been randomized by a small amount in order to avoid
  clutter.  Due to lack of colors in some cases, not all of the MS
  secondaries can be shown.  Note in the upper panel the pair composed
  of a turnoff/subgiant primary and a WD secondary (HIP 18824/NLTT
  12412, with the WD located at M$_{\rm V} \sim 14$ and $V-J \sim 1$).
  Even though there is no FGK MS star in this pair, we keep it in our
  program in case the WD is of DA type in order to compare the age of
  the same system obtained from two different methods. }.
\label{fig:sel2}
\end{figure}

We proceed to identify systems with evolved components based on the
comparison with the well known {\it Hipparcos} CMD obtained with $B-V$
colors and $V-$band absolute magnitudes. This is illustrated in
Figures \ref{fig:sel1} and \ref{fig:sel2}, corresponding to the two
distinct categories we defined above according to the existence of
independent {\it Hipparcos} measurements for both or just one of the
binary components, respectively. The small background dots in the
lower plots are {\it Hipparcos} stars within 100 pc of the Sun and
with better than $5\sigma$ parallaxes, obtained from the Vizier
Service\footnote{http://vizier.u-strasbg.fr/viz-bin/VizieR}. They
clearly show the population of field stars we are after, i.e., evolved
stars leaving the MS on their way to become red giants. The top panels
of these two figures show a CMD based on the $V-J$ color, which is the
reference color in the GC04 and LB07 catalogs. The small dots in the
$V-J$ CMDs are a subset of the LSPM-North proper-motion catalog that
includes stars with good distances within 33 pc from the Sun
\citep{lepine05}. Since not listed in GC04 and LB07, optical $B-V$
colors for all stars in the catalogs were also obtained from Vizier
queries to the {\it Hipparcos} catalog.

Wide binaries with recently evolved primaries were selected by
defining the area contained within the dashed lines in the lower CMDs
of Figures \ref{fig:sel1} and \ref{fig:sel2}.  This area was designed
so that it encompasses the largest fraction possible of turnoff stars
and subgiants, and at the same time attempting to avoid serious
contamination from MS stars on the blue side.  Given the smaller
density of stars on the red side, we were less restrictive on that end
and extended the selection box so that it includes the base of the red
giant branch.

From the pool of wide binaries in GC04 and LB07 with both components
in {\it Hipparcos}, we selected all pairs for which any of the two
stars fall within the search area defined above, finding 6 with MS
components bluer than $B-V \sim 1.40$, approximately the boundary
between K and M dwarfs. Since our source catalogs are based on the
original {\it Hipparcos} reduction \citep{ESA97}, we performed this
initial search using parallax measurements from that database.
Repeating the same exercise using the parallax measurements from the
revised {\it Hipparcos} reduction \citep{hipparcos07} produced two
more pairs, which we included in the sample. An additional pair, HIP
10305/HIP 10303, long known to be a wide binary system (e.g.,
\citealt{vdb58}), was part of a sample on a project different from
that described in this paper, but, finding it suitable for
gyrochronology, it was included in our sample.  The 9 pairs selected
in this way are shown in Figure \ref{fig:sel1} and listed in Table
\ref{tab:hipp_doubles}. The parallaxes reported in this Table and
throughout this work are from the 2007 {\it Hipparcos} reduction.

The second pool of wide binaries from GC04 and LB07 are those with
only one of their components in {\it Hipparcos}. The selection of
systems with a recently evolved component from this group is performed
exactly as done for the first group above, only that this time the
secondary is assigned the parallax measured for the primary.  Since we
could not find $B-V$ measurements for all the secondaries in this
group, we only required for these stars that $V-J < 3.0$, which
approximately marks the boundary between K and M dwarfs. The 10
targets found from this group are illustrated in Figure \ref{fig:sel2}
and listed in Table \ref{tab:hipp_nonhipp}. Note that one of these
pairs is composed of a turnoff/subgiant primary and a WD secondary.
While not useful for gyrochronology purposes, we decide to include it
in our program because it may at some point serve us to compare the
age of the system as derived from two different methods.

As will be seen in \S\,\ref{ivan}, the radial velocities and
metallicities that we derive for an initial subsample of the targets
selected in this section confirms that these are indeed binary
systems. The mean [Fe/H] difference is $0.013 \pm 0.084$, i.e.
consistent with zero, and the radial velocities agree within $\sim 2$
km/s (the mean difference for 10 pairs is $-1.05 \pm 1.76$ km/s). The
exception are three pairs showing radial velocity differences of $3-4$
km/s (HIP 94076/HIP 94075, HIP 115126/NLTT 56465, and HIP 114855/NLTT
56278). For truly bound wide binary sytems, such large differences
between the radial velocities of their components can only be
explained by detected orbital motion (see figure 1 in \citealt{cg04}),
or else if one of the components has a close, undetected companion
that induces periodic radial velocity variations. Indeed, this is the
case of HIP 94076, a known spectroscopic binary, and we plan to obtain
new epochs of spectroscopy for the other two cases.

\subsection{Wide binaries with white dwarf components}\label{sel:WDs}

The CG04 catalog contains 82 pairs with a WD component, and GC04
contains 20 pairs composed of a WD and an {\it Hipparcos} star. These
numbers, however, include all types of WDs as well as all types of MS
components, while we are interested here only in pairs composed of a
DA WD and a MS star of FGK type.

A first selection and age-dating of pairs from CG04 and GC04
containing WD components of DA type was recently performed by
\citet{garces11}, who took advantage of such systems in order to
obtain the ages of a number of GKM stars aimed to calibrate the time
evolution of high-energy emissions associated to stellar activity in
low-mass stars.  Although the majority of wide binaries in their
sample contain M dwarf components, there are at least 7 systems with
GK components that could be useful for the purposes of gyrochronology.
We list these in Table \ref{tab:WDs} along with relevant photometric,
astrometric, and kinematic data, as well as the ages derived by their
work.

Additionally, we obtain 11 more pairs composed of a WD plus a GK main
sequence star from \citet{zhao11}, who investigated the time evolution
of chromospheric activity levels in solar-type dwarfs belonging to
wide binaries selected from the original Luyten proper motion survey
\citep{luytenWDs} and from \citet{giclas71}.  Relevant data for these
pairs are also listed in Table \ref{tab:WDs}, when available.

We also inspected the 21 pairs with WD components in the SLoWPoKES
catalog \citep{dhital10}.  However, all stars in this sample are faint
(both primaries and secondaries in the range $r = 16 - 20$ mag) and,
moreover, all the MS companions of these WDs turned out to be M
dwarfs.

\section{Stellar Parameters for an Initial Sample}\label{ivan}

As a first step in our gyrochronology program, we set out to determine
the ages of a number of the wide binary systems selected in
\S\,\ref{targets}. The sample of wide binaries with WD components of
\S\,\ref{sel:WDs} has already been studied by \citet{garces11} and
\citet{zhao11}, and thus we simply take their resulting ages and list
them in Table \ref{tab:WDs} along with all the relevant data for these
systems. In what follows, we concentrate on the samples of wide
binaries with turnoff or subgiant components.

Since the ages of recently evolved stars are derived with the use of
theoretical isochrones, we first need to obtain sufficiently precise
measurements of the stellar parameters ($T_{\rm eff}$, [Fe/H], and
$\log g$) of the evolved component in these targets.  In principle,
this is only necessary for the component of the binary that is best
positioned to provide information on the age of the
system\footnote{While evolved stars have typically been the best type
  of star for this, we note that today very high-precision stellar
  parameters can be derived for solar analog stars, and those can in
  turn be used to derive reasonably good isochrone ages, even for
  unevolved stars on the main-sequence (e.g.,
  \citealt{bau10}).}. Due to a number of reasons, however, we attempt
to obtain isochronal ages for both components of our pairs, when
possible. First, the procedure of age-dating via isochrones not only
provides ages but can also be used to better characterize the stellar
properties and evolutionary state of the stars being studied. This is
important in our context because the entire idea behind gyrochronology
makes sense only for stars on the MS, whose rotation periods respond
to their spin-down due to stellar activity. Therefore, we need to rule
out as best as we can the possibility that some of the MS secondaries
of the wide binaries selected in \S\,\ref{targets} may have started
themselves to evolve away from the MS, at which point the physical
expansion of the star would affect its surface rotation rate and thus
invalidate the system as a useful constraint for gyrochronology.
Second, even for the cases where the MS components of the selected
binaries are indeed FGK stars on the MS, we want to explore the
possibility of obtaining a better constraint on the age of the system
by forcing the procedure of isochrone fitting to consider the two
stars as coeval. Third, for the cases where both stars are able to
provide independent isochronal ages for the system, we would like to
explore closely those pairs that produce inconsistent results, if any.
Fourth, independent determinations of the metallicities of both
components of our wide binaries automatically provide us with a check
on our measurement errors, and may also serve as an additional test
for the binary nature of our pairs\footnote{In
  the context of the scenario of \citet{kou10} for the formation of
  wide binaries, there is the possibility that the components of a
  genuine, gravitationally bound system may truly have slightly
  different metallicities, which would correspond to metallicity
  variations or gradients across the parent young star cluster.
  Additionally, it is becoming increasingly more clear that small
  abundance changes could be produced by planet formation
  \citep{mel09,ivan10}, and small differences between stars in binary
  systems have been detected too \citep{desidera04,ramirez11}.  Thus we
  expect the metallicities of the two stars in a binary system to be
  similar but not necessarily identical}.  Therefore, for the 
measurement of stellar parameters prior to age determination, we target 
both components of our selected wide binaries.

\subsection{Observations and Data Reduction}

Most of our target stars accessible from the Southern hemisphere were
observed with the MIKE spectrograph \citep{bernstein03} on the 6.5\,m
Clay Telescope at Las Campanas Observatory on 2010 September 21-22 and
2011 January 4. We used a narrow slit ($0.35\arcsec$), which delivers
data with spectral resolution $R=\lambda/\Delta\lambda\simeq65\,000$
(at $\lambda\simeq6000$\,\AA) and the standard setup that allows
complete wavelength coverage in the 3400--9100\,\AA\ spectral
window. These spectra were reduced using the CarnegiePython
pipeline,\footnote{http://obs.carnegiescience.edu/Code/mike} which
employs multiple bias and flat-field frames to correct for
instrumental imperfections and ThAr lamp exposures taken throughout
each night for wavelength calibration, in addition to 
co-adding multiple exposures of the
same object. The signal-to-noise ratio (S/N) of our reduced spectra
(per pixel) varies between about 100 and 600 at $\lambda=6000$\,\AA\
with a median of about S/N=400 (at $\lambda=4000$\,\AA\ the median S/N
is about 150).

Reflected sun-light spectra from the asteroid Hebe were acquired on
2010 September 21 for solar reference. Observationally, asteroids
behave like point sources, thus making their data acquisition and
reduction identical as that for the rest of our targets. This is not
the case of scattered sky-light or Moon observations, which are
sometimes also used for solar reference. The use of asteroid
observations allows a more precise differential analysis.

The radial velocities of the stars were estimated from the Doppler
shifts of spectral line cores of hundreds of \fei\ features. We used
the rest wavelengths measured in the laboratory by \cite{nave94} and
measured core wavelengths in the observed spectra by fitting a
parabola to the 7 data points closest to the flux minimum of each
line. The internal precision of our radial velocity measurements is
about 0.35\,\kms. However, systematic errors due to core wavelength
shifts produced by surface convection are of order 0.5\,\kms\
\citep[e.g.,][]{gray09,ramirez09:iau}. The use of cross-correlation
with radial velocity templates would not remove entirely this error
because of the wide range of stellar parameters of our sample and the
fact that the impact of granulation is still poorly known even for
standard stars. Nevertheless, the accuracy of our radial velocities
($\simeq0.5$\,\kms) is sufficient for our purposes. Our observing log
and radial velocities derived are listed in Table~\ref{t:obslog}.

\subsection{Spectroscopic Analysis}

The fundamental atmospheric parameters $\teff$, $\logg$, and $\feh$ of
a star can be estimated using a variety of techniques.\footnote{We use
  the standard notation: $\feh=\afe-\afe^\odot$, where
  $\afe=\log(N_\mathrm{Fe}/N_\mathrm{H})+12$ and $N_\mathrm{X}$ is the
  number density of X atoms in the stellar photosphere.} In addition
to employing only the observed spectra, photometric data as well as
trigonometric parallaxes can be used to constrain one or more of these
quantities. Here we describe the techniques used in our work. The
atmospheric models adopted are from the MARCS grid of standard
chemical composition
\citep{gustafsson08}.\footnote{http://marcs.astro.uu.se} The
curve-of-growth analysis was made using the 2010 version of the
spectrum synthesis code MOOG
\cite[e.g.,][]{sneden73}.\footnote{http://www.as.utexas.edu/$\sim$chris/moog.html}

\begin{figure}
\epsscale{0.7}
\plotone{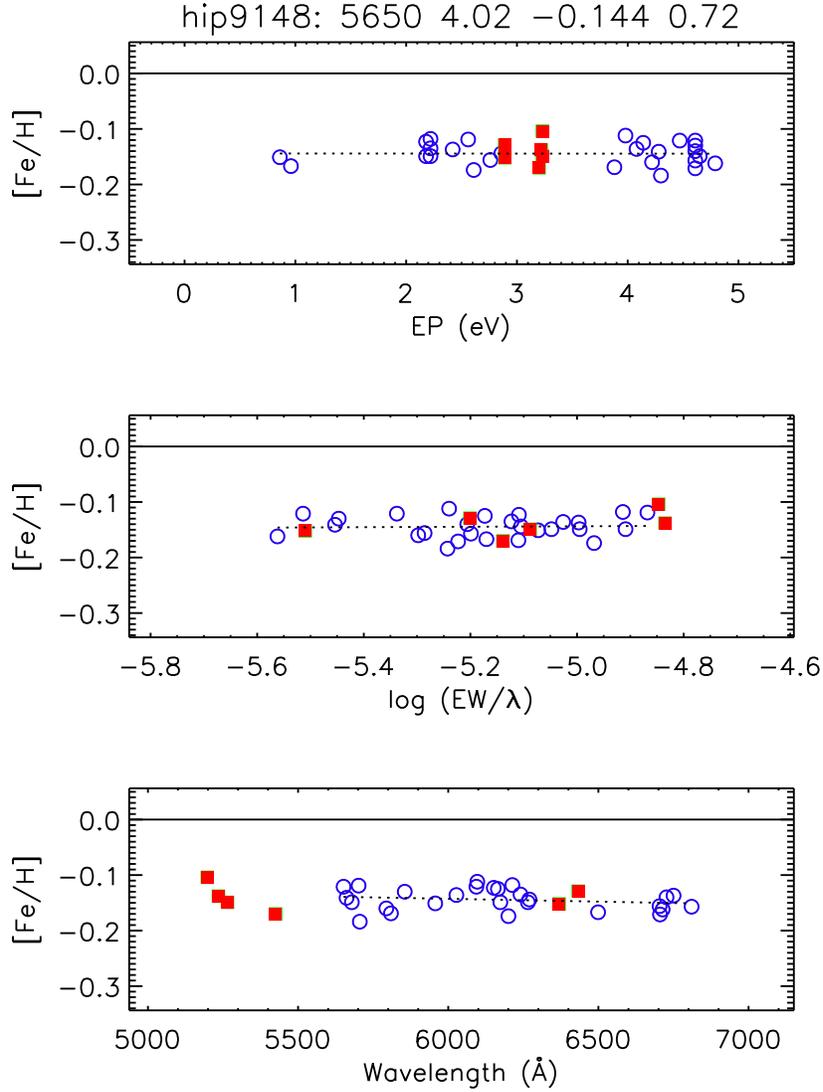}
\caption{Measured iron abundance as a function of excitation potential (top panel), reduced equivalent width (middle panel), and wavelength (bottom panel) for HIP\,9148. Open circles (filled squares) correspond to \fei\ (\feii) lines. Dotted lines are linear fits to the data. The legend on the top panel shows the star name followed by the derived parameters: $\teff$, $\logg$, $\feh$, and $v_t$. The solid line is at solar metallicity.}
\label{f:sp}
\end{figure}

We started by using the standard iron line spectroscopic approach,
which forces excitation and ionization balance of iron lines. A first
guess of the parameters is made and iron abundances are computed for a
number of neutral (\fei) and singly ionized (\feii) iron lines. The
parameters are then fine-tuned to remove any correlation between iron
abundance and excitation potential (EP) of \fei\ lines (therefore
forcing excitation balance) and to minimize the difference between the
mean iron abundances inferred from \fei\ and \feii\ lines separately
(thus achieving ionization balance). Simultaneously, the correlation
between \fei\ abundance and line strength is controlled (i.e., the
correlation is minimized) with the microturbulent velocity parameter
($v_t$). Fig.~\ref{f:sp} shows an example of the end product of this
procedure. The iron line-list adopted (including atomic data) is from
\cite{asplund09:review}, who made a careful selection of unblended
features for their solar abundance analysis. The abundances used in
this procedure are differential, on a line-by-line basis, using the
solar abundances inferred from our solar (Hebe) spectrum as
reference. We adopted $v_t^\odot=1.0$\,\kms, although the exact value
of $v_t^\odot$ has a minor impact on our results.

Errors in the derived parameters are estimated from the uncertainty in
the abundance versus EP slope (for $\teff$) and the line-to-line
scatter of the mean \fei\ and \feii\ abundances (for $\logg$).  Since
we force the EP slope to be zero, a slightly positive (negative) EP
slope implies a $\teff$ too low (high) by a certain amount. We use the
$\Delta \teff$ amount that corresponds to an EP slope of $\pm 1
\sigma$, where $\sigma$ is the error of the zero slope when using the
adopted $\teff$. For the error in $\logg$, we consider the maximum and
minimum $\logg$ values such that the mean \fei\ $-$ \feii\ abundance
difference is consistent with zero within the $1\sigma$ line-to-line
scatter as the upper and lower limits of the derived $\logg$.

A second set of parameters was obtained using colors to derive
$\teff$. The recent metallicity-dependent color calibrations by
\cite{casagrande10} for the following color indices: $B-V$, $b-y$,
$V-J$, $V-H$, $V-K_s$, and $J-K_s$, were used. Observed magnitudes and
colors of our sample stars were taken from the General Catalogue of
Photometric Data
\citep{mermilliod97}\footnote{http://obswww.unige.ch/gcpd/gcpd.html}
and the 2MASS and Hipparcos/Tycho catalogs. We made sure that the
adopted photometry was not blended (i.e., we excluded mainly old
measurements for the unresolved systems) and avoided uncertain 2MASS
photometry for the brightest stars. We did not apply the
\cite{casagrande10} formulas to giant stars because their work is
restricted to dwarf and subgiant stars. Errors in the photometry and
the color-to-color $\teff$ scatter provides us with an estimate of the
error in $\teff$.  Photometric errors are simply propagated into the
color$-\teff$ relations to obtain the error in $\teff$ for a given
color. Then these errors are used as weights when computing the final
$\teff$ value. The $\teff$ error is obtained using the formula for the
sample variance, again using the $\teff$ errors for each color as
weights.

Given a photometric $\teff$, the surface gravity was determined using
two methods. The first one is the same as in the iron line analysis,
i.e., $\logg$ is fine-tuned so that the mean iron abundances from
\fei\ and \feii\ lines agree (ionization balance). In the second case
we determine $\logg$ using the stars' trigonometric parallax from the
new reduction of Hipparcos data \citep{hipparcos07} and theoretical
isochrones. This method is described in detail in \S\,\ref{ages}. Note
that in this latter case it is not guaranteed that the iron abundances
inferred from \fei\ and \feii\ lines are the same. Moreover, either if
$\logg$ is inferred forcing ionization balance or using isochrones,
the \fei\ abundances will in general show a correlation with EP. Thus,
the line-to-line scatter of the iron abundances inferred using
photometric temperatures will be larger than that obtained by forcing
excitation and ionization balance of iron lines. However, this does
not imply a superiority of one method over another; it simply reflects
the nature of the different approaches to measure the stellar
parameters.

\begin{figure}
\epsscale{0.8}
\plotone{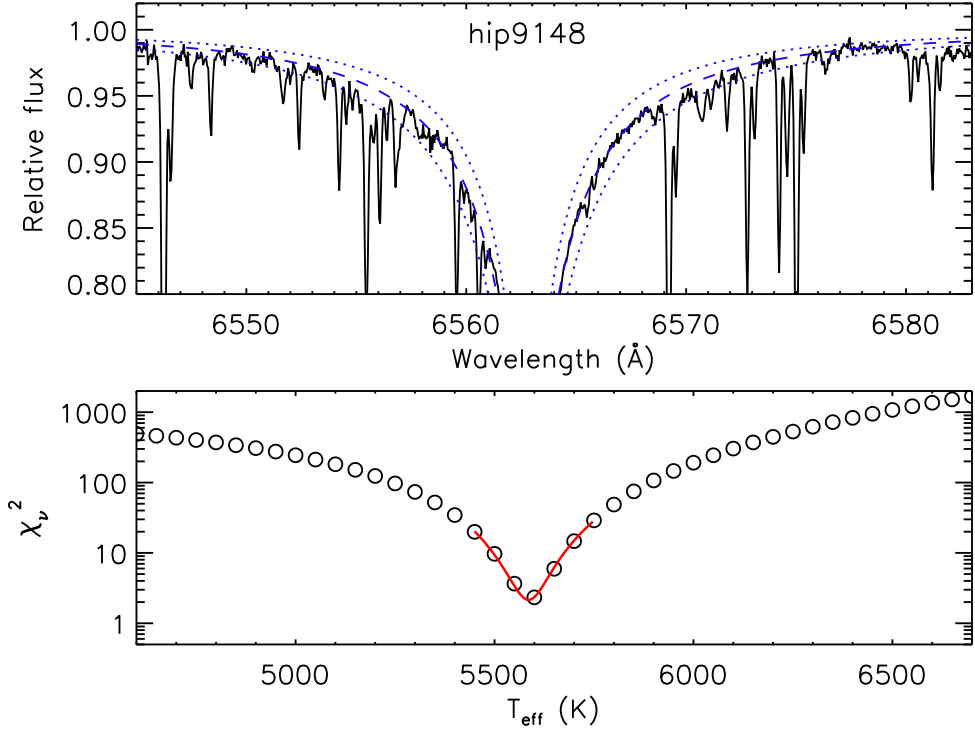}
\caption{Top panel: Observed (solid line) and theoretical (dashed and dotted) \ha\ line profiles for HIP\,9148. The dashed line corresponds to the best fit $\teff$ \ha\ model while dotted lines show profiles for $\teff\pm200$\,K. Bottom panel: reduced $\chi^2$ of observation minus model differences as a function of model $\teff$ (open circles). Only the regions not affected by narrow spectral lines are used in this computation. The solid line is a parabolic fit to the 7 points closest to the minimum, which allows a better $\chi^2$ minimization.}
\label{f:halpha}
\end{figure}

Finally, a third estimate of effective temperature can be obtained by
analyzing the wings of Balmer lines, in particular \ha. The depth of
these wings is highly sensitive to $\teff$ and if the other stellar
parameters can be constrained independently, very precise $\teff$
values can be inferred from a $\chi^2$ minimization of observation
minus theoretical models of \ha\ line profiles
\cite[e.g.,][]{barklem02,ramirez06,ramirez11}. A proper continuum
normalization is required for this method to provide accurate
effective temperatures. Our \ha\ line profiles were normalized taking
advantage of the smooth variation of the blaze function across
spectral orders. Polynomial fits were used to trace the upper
envelopes of spectral orders above and below the order containing the
\ha\ line. They were then interpolated to trace the continuum of the
\ha\ order. We used a grid of theoretical \ha\ line profiles computed
by \cite{barklem02}, which is based on MARCS atmospheric models and
the self-broadening theory developed in
\cite{barklem00}.\footnote{This grid of theoretical \ha\ line profiles
  is available online at http://www.astro.uu.se/$\sim$barklem.} A
$\chi^2$ minimization routine allowed us to find the best model fits
to our data and therefore to estimate $\teff$ and its associated
error. Fig.~\ref{f:halpha} illustrates this technique. We obtained
$\teff=5741\pm40$\,K for our solar (Hebe) spectrum, in very good
agreement with the solar $\teff$ inferred by \cite{barklem02} using
the \ha\ line from the very high quality solar spectrum by
\cite{kurucz84}. The fact that the \ha\ temperature of the Sun does
not perfectly agree with the nominal value of 5777\,K suggests that
there is still room for improvement in the modeling of Balmer
lines. We increased the $\teff$ values derived from our \ha\ analysis
by 36\,K as a first order correction, given that this brings the solar
$\teff$ up to its expected value. Similar to the case of photometric
$\teff$, once the effective temperature was determined from the \ha\
line analysis, we derived $\logg$ values using ionization balance as
well as isochrones.

Note that all the methods described above require a previous knowledge
of the atmospheric parameters, which are the quantities we want to
derive. Thus, iterative procedures are necessary to obtain a final,
self-consistent solution for the $\teff,\logg,\feh$ set. We repeated
all calculations as many times as necessary to guarantee that no more
iterations are required to improve these solutions.

The $\teff$, $\logg$, $\feh$ values derived using the various
techniques described in this Section are listed in Table~\ref{t:pars}.

\subsection{Adopted Parameters}\label{adopted}

All methods of atmospheric parameter determination have
limitations. For example, the excitation/ionization balance of iron
lines assumes local thermodynamic equilibrium (LTE), which is a useful
but not necessarily correct assumption
\citep[e.g.,][]{asplund05:review}. Photometric temperature scales have
uncertain zero points, and even though they seem to agree well with
$\teff$ scales based on interferometric observations of stellar
angular diameter, the latter are typically based on the analysis of
giant stars, whose surfaces are far from being static and
well-defined, as it is assumed in the measurement of angular diameters
\cite[e.g.,][]{koesterke08,chiavassa10}. Also, the modeling of \ha\
line profiles relies heavily on the theory of self-broadening, and
different prescriptions may lead to significantly different results,
as explored in detail by \cite{barklem02}. Surface gravity
determinations based on ionization balance rely also on the assumption
of LTE while the isochrone approach depends on stellar structure and
evolution calculations and may be affected by statistical biases
\cite[e.g.,][]{pont04,jorgensen05}.

It is therefore not surprising to find differences in the stellar
parameters derived with different methods for the same object in
Table~\ref{t:pars}. Our sample size is too small and it covers too
wide of a region of stellar parameter space for us to clearly uncover
systematic differences between the different methods. In any case,
discrepant values typically point to systematic errors affecting
differently each technique whereas values in good agreement suggest
that the impact of these errors is relatively small. Thus, by taking
the average value of all the measurements reported in
Table~\ref{t:pars} for each object we improve upon values that are
only slightly affected by systematic errors while assigning realistic
error bars to stars for which their derived parameters likely suffer
from severe systematic errors.

Our adopted parameters are obtained as the weighted average of all
measurements available to each star. However, effective temperature
errors lower than 35\,K were set to 35\,K before averaging. Similarly,
the lowest adopted $\logg$ and $\feh$ errors were 0.03 and 0.04\,dex,
respectively. This was done to avoid giving too much weight to a
particular value. Small internal errors can in many cases be the
result of numerical artifacts that do not reflect the true errors of
the measurement. By adopting reasonable minimum errors we obtain more
reliable, and therefore accurate, average values.

The errors adopted are the linear sum of the standard error and the
sample variance. The first term takes into account the random errors
of each measurement assuming that there are no systematic differences
between the various values. The sample variance, on the other hand, is
sensitive to systematic differences. Thus, by adding the sample
variance to the standard error a more realistic error bar is obtained.

Our adopted parameters and their associated errors, computed as
described in this Section, are given in Table~\ref{t:final_pars}.

\section{Stellar ages}\label{ages}

We compute the ages of our initial sample of recently evolved wide
binaries by making use of theoretical isochrones.  In this technique,
the star under study is placed on a CMD and its location compared to
theoretical predictions of stellar evolution.  Isochrone points close
to the observed stellar parameters are then used to derive the mass
and age of the star (e.g., \citealt{lachaume99}).  While the location
of any star on the CMD is determined just by its absolute luminosity
$M_{\rm V}$ and surface temperature $T_{\rm eff}$ (or color), many
combinations of stellar parameters ($\log g$, [Fe/H], mass, and age)
can go through that point, so that in order to infer the age of the
star some of these parameters must also be given.  Typically, the
fundamental atmospheric parameters $T_{\rm eff}$, $\log g$, and [Fe/H]
are measured from the star's spectrum, photometry, or a combination of
both (\S\,\ref{ivan}), and they are enough to determine both the age
and mass of the star via isochrones.  In our case, given that for some
stars different methods to estimate $\log g$ produced discrepant
answers (\S\,\ref{adopted}), we avoid this parameter and instead use
the luminosity of the star, $M_{\rm V}$ (i.e., determined by its
apparent magnitude and parallax).

In order to determine which isochrone fits best any particular star,
we build the probability distribution of the stellar age by computing
the likelihood that any given isochrone passes near the corresponding
set of stellar parameters, accounting for the uncertainties on those
parameters. The age probability distribution is also called the ``G
function'', and the procedure is widely used nowadays (e.g.,
\citealt{lachaume99,reddy03,nordstrom04,allende04,pont04}). Assuming
the errors in our stellar parameters ($\sigma_{T_{\rm eff}}$,
$\sigma_{\rm [Fe/H]}$, and $\sigma_{M_{\rm V}}$) have Gaussian
probability distributions, the likelihood for a point in a given
isochrone can be written as

\begin{equation}\label{eq:likelihood}
P(T_{\rm eff},{\rm [Fe/H]},M_{\rm V}) \propto
\exp \left[ -(\Delta T_{\rm eff})^2\over 2\sigma_{T_{\rm eff}}^2 \right]
\exp \left[ -(\Delta [{\rm Fe/H}])^2\over 2\sigma_{[{\rm Fe/H}]}^2 \right]
\exp \left[ -(\Delta M_{\rm V})^2\over 2\sigma_{M_{\rm V}}^2 \right],
\end{equation}

\noindent where $\Delta T_{\rm eff}$, $\Delta [{\rm Fe/H}]$, and
$\Delta M_{\rm V}$ are the differences between the measured stellar
parameters and those corresponding to the point in the isochrone under
consideration. The integral of this likelihood over all the parameter
space of the set of isochrones then gives the age probability
distribution

\begin{equation}\label{eq:age}
P({\rm age}) = \int P(T_{\rm eff},{\rm [Fe/H]},M_{\rm V})\, {\rm d}T_{\rm eff}\,
{\rm d[Fe/H]}\, {\rm d}M_{\rm V}.
\end{equation}

\noindent In practice, we only integrate over the volume defined by
three times the $1\sigma$ uncertainties from the measured stellar
parameters, which we verified already accounts for most of the
contribution to the probability distribution from the entire set of
isochrones. Being the most likely value, we then adopt the peak of the
G-function as the age of the star.

The choice on how to determine the exact location of the peak of the
G-function has a small effect on the adopted age, especially given the
size of our $1\sigma$ errors, typically larger than $\sim$ 0.5 Gyr.
Therefore, we do not implement at this point anything more
sophisticated than just adopting as our age the center of the bin
where the peak occurs, and leave for a later stage in this program any
refinement, if at all needed. Due to the discrete nature of the
procedure of isochrone fitting, this lack of ``smoothing'' of the
G-functions leads to the appearance of spurious structure in the
distribution of ages of large samples of stars (e.g.,
\citealt{nordstrom04}), so a more sophisticated approach to determine
the exact location of the peak is relevant for population studies,
which is not the case here.

\begin{figure}
\epsscale{1.0}
\plotone{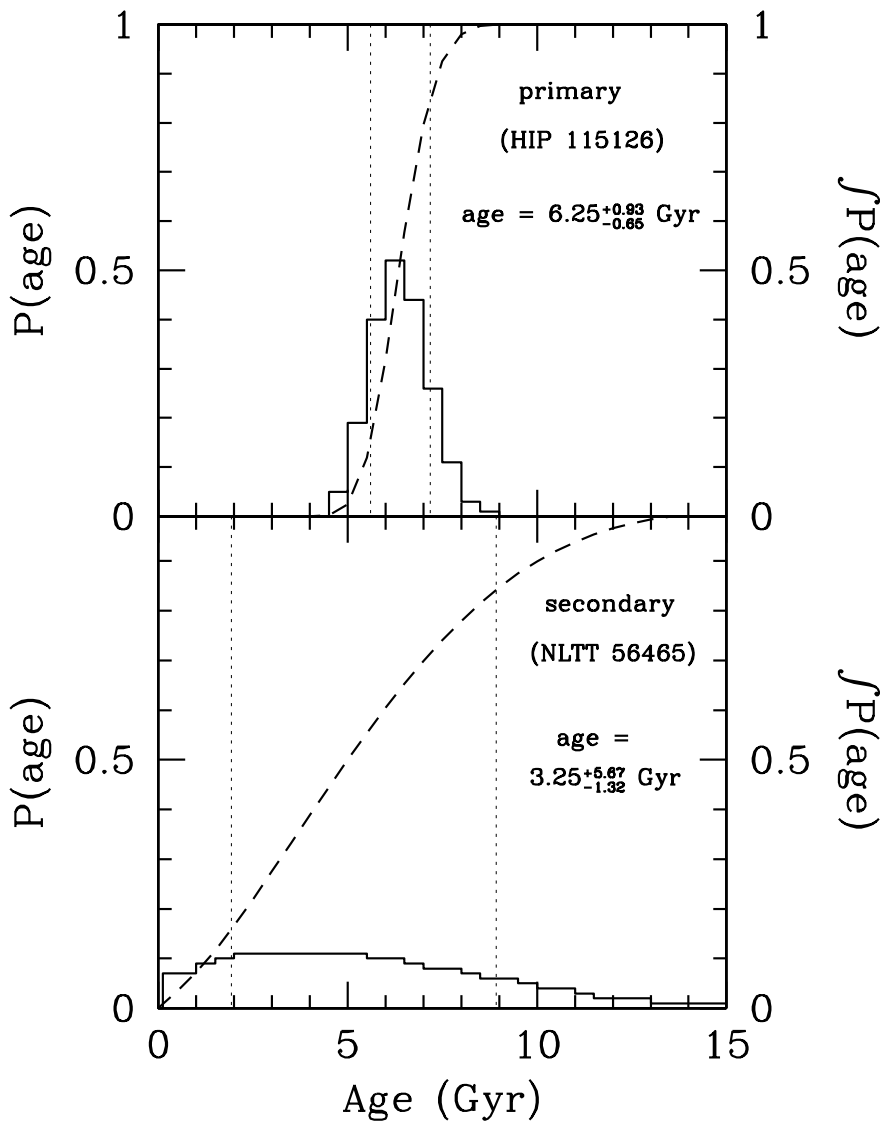}
\caption{Age probability distributions (solid lines) for the
  components of a representative case (not the best, not the worst) of
  our initial sample of wide binaries with recently evolved
  components. The vertical dotted lines indicate the location of the
  adopted $1\sigma$ errors to both sides of the peak of the
  distributions, obtained from the cumulative function of the age
  distribution (dashed lines). The evolved primary provides a
  significantly better constraint on the age than the MS secondary.}
\label{fig:singles}
\end{figure}

As an illustration, Figure \ref{fig:singles} shows the resulting age
distributions and $1\sigma$ uncertainties around the most likely age
for the pair HIP 115126/NLTT 56465 (solid lines). We normalize the
distributions so that the area below is equal to 1. The adopted errors
are computed from the cumulative function of the age probability
distribution (dashed lines), assuming the latter is well approximated
by a Gaussian. However, as is clear from the bottom panel in Figure
\ref{fig:singles}, this G-function is not always close to a Gaussian,
and thus we adopt different $1\sigma$ errors to both sides of the
peak. Our $1\sigma$ lower and upper limits mark the age interval of
cumulative probability between 16\% and 84\%.

Using this procedure we compute the ages of all stars (both primaries
and secondaries) in our initial sample. We use Yale-Yonsei isochrones
\citep{yi01,demarque04} sampled with constant steps of 0.1 Gyr in age
and 0.02 dex in [Fe/H], covering ages from 0.1 to 15 Gyr. In
constructing the age probability distribution, we experimented with
different bin sizes and found that 0.5 Gyr/bin was an optimal choice
and smaller bins did not change the resulting ages. For the case of
binaries with both components in {\it Hipparcos}, we adopt for both
stars the primary's parallax, which for all pairs in this category is
the one with the better measurement (i.e., smaller error). The
resulting ages and uncertainties for all stars that could be observed
with enough signal-to-noise are listed in Table \ref{tab:ages}

Following the same steps leading to the derivation of the isochronal
age via equations \ref{eq:likelihood} and \ref{eq:age}, one can derive
an isochronal $\log g$ through the construction of the gravity
probability distribution. Our results for this isochronal $\log g$ are
listed in Table \ref{t:final_pars} (labeled as $\logg_{\rm iso}$),
where they can be compared to the average of the resulting $\log g$
derived from all other methods. Since the $\log g_{\rm iso}$ are
obtained from our final $T_{\rm eff}$ and [Fe/H] for each star, as
listed also in Table \ref{t:final_pars}, they are our preferred
values, above the $\log g$ obtained as the average of different
methods in Section \ref{ivan}. 

\begin{figure}
\epsscale{1.0}
\plotone{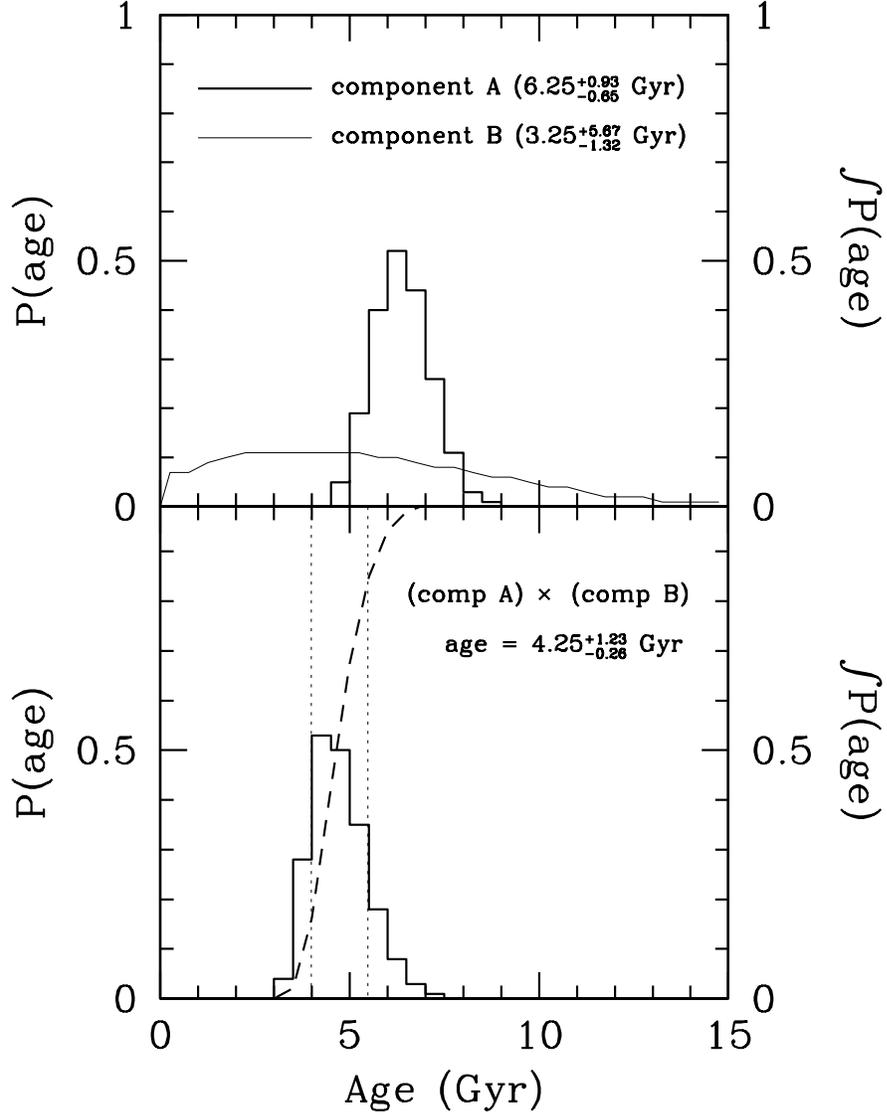}
\caption{Age probability distribution assuming coevality for the pair
  shown in Figure \ref{fig:singles}, i.e., considering the two
  components simultaneously (see text in \S\,\ref{ages}). {\it Upper
    panel}: independent, single-star age probability distributions for
  HIP 115126 (A, primary) and NLTT 56465 (B, secondary). {\it Lower
    panel}: Joint probability distribution (solid line), obtained by forcing the
  isochrone fitting to pass through the two components of the binary
  using equation \ref{eq:joint}.}
\label{fig:bothstars}
\end{figure}

Next we consider the interesting possibility of improving the ages of
those systems in our sample for which the two components were able to
provide an independent age. In these cases, since both components
contain some information on the age of the system, instead of treating
them independently we can treat them simultaneously by forcing the
procedure of isochrone fitting to consider the two stars as coeval.
This can be done by computing the age probability distribution of the
binary system, which would be given by

\begin{equation}\label{eq:joint}
P({\rm age}) = \int P_{\rm A}\,P_{\rm B}\, {\rm d}T_{\rm eff}\, {\rm d[Fe/H]}\, {\rm d}M_{\rm V},
\end{equation}

\noindent where $P_{\rm A}$ and $P_{\rm B}$ are the likelihoods for
components A and B, respectively, and given by equation
\ref{eq:likelihood}. While equation \ref{eq:joint} gives the same
weight to both stars, one could assign different weights based on a
number of criteria, but we choose to not attempt this. Figure
\ref{fig:bothstars} illustrates the case of HIP 115126/NLTT 56465, the
same pair shown in Figure \ref{fig:singles}. The net effect of
considering the two components simultaneously was to shift the entire
G-function of the evolved primary towards younger ages, closer to the
broad peak of the secondary's G-function.  The nominal precision of
the resulting combined age (i.e., the size of the $1\sigma$ region
around the peak), however, remains esentially the same as that for the
evolved primary alone.

All the ages derived up to this point come directly from the adopted
set of isochrones and the fitting procedure detailed above. The use of
a different set of isochrones will, in principle, change our results.
Moreover, isochrone ages are known to suffer from a number of
systematic biases arising, for example, from the fact that, due to the
different timescales characteristic of different evolutionary stages,
isochrones do not evenly map the relevant parameter space of stellar
parameters. There is a significant literature on the subject of such
age biases and the available ways to correct for them
\citep{pont04,nordstrom04,jorgensen05}. These systematics are
important when dealing with large samples of stars and attempting to
extract conclusions of a statistical nature, such as those related to
the age-metallicity relation in the solar neighborhood (e.g.,
\citealt{casagrande11}). Since the main purpose of the present paper
is to introduce our program for gyrochronology with wide binaries and
demonstrate that precise ages (i.e., with reasonably small internal
errors) can be achieved for them, any refining of the ages derived in
this section will be attempted later, if at all necessary, and thus we
do not implement any corrections for statistical biases or derive ages
from other isochrone grids at this point.

Nevertheless, in order to provide readers with a quantitative idea of
the differences expected when accounting for statistical biases and
isochrone grid choice, we compute the ages obtained by making use of
the ``web interface for the Bayesian estimation of stellar
parameters'', maintained at the Osservatorio Astronomico di
Padova\footnote{http://stev.oapd.inaf.it/cgi-bin/param} and described
in \citet{dasilva06}. This publicly available tool uses a different
set of isochrones than the one used in this paper (namely, that by
\citealt{girardi00}) and accounts for some of the biases discussed
above. The tool requires the selection of priors for the Bayesian
analysis, for which we choose an initial mass function from
\citet{chabrier01}, and a constant star formation rate in the interval
from 1 to 12 Gyr. The resulting ages and their uncertainties are
listed in Table \ref{tab:ages} next to our own determinations.

\begin{figure}
\epsscale{0.6}
\plotone{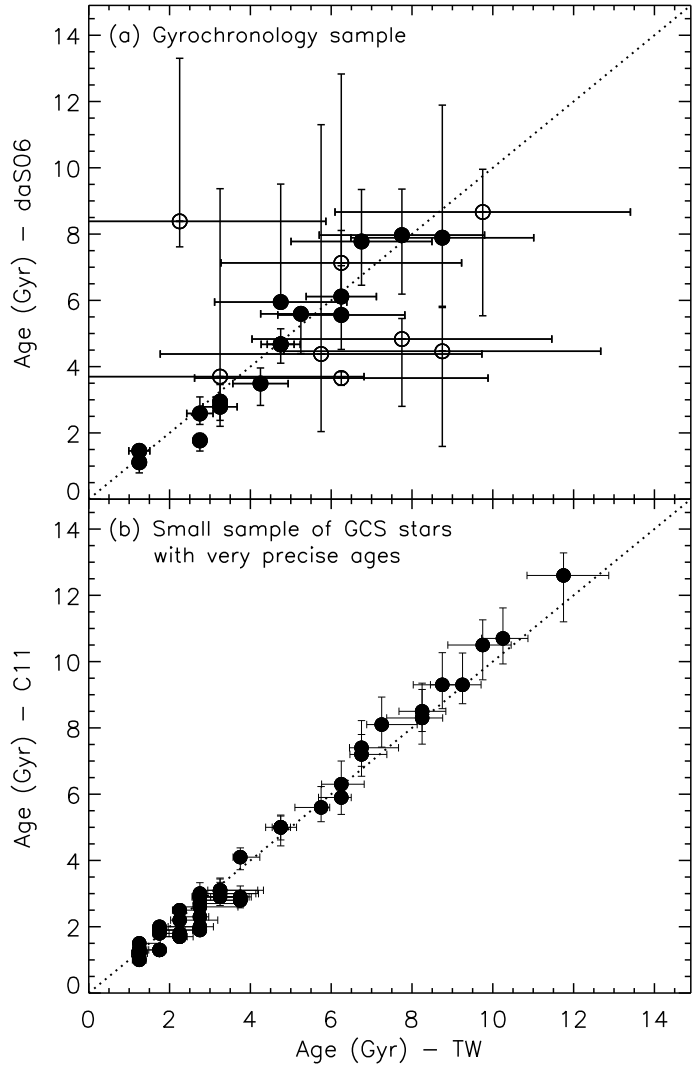}
\caption{(a) Direct comparison between the isochronal ages derived in
  this work and those obtained using the \citet{dasilva06}
  prescription, which accounts for statistical biases and makes use of
  a different set of isochrones. The filled circles are stars
  considered robust by \citet{dasilva06} (age/$\sigma_{\rm age} > 3$).
  The agreement is excellent, with only $3-4$ significantly discrepant
  cases out of a total of 24 stars. (b) Comparison of ages from the
  GCS survey for a small sample of GCS stars with very well determined
  ages, as derived by \citet{casagrande11} and our ages, obtained
  using the exact
  same input data from C11.}
\label{fig:dasilvagcs}
\end{figure}

In Figure \ref{fig:dasilvagcs}a we show the comparison between our
isochronal ages (for the components of the wide binaries treated
independently) and those obtained accounting for statistical biases
following the \citet{dasilva06} prescription and a different set of
isochrones. In most cases the ages are remarkably similar, even in the
adopted uncertainties. The mean difference of the maximum likelihood
ages, for the cases where the \citet{dasilva06} ages are at least
three times larger than their $1\sigma$ errors, is only $0.1\pm0.6$
Gyr, thus giving reliability to our isochronal ages and suggesting
that inclusion of the statistical biases is not crucial for our
purposes.

We also tested our age determination method against the Bayesian
implementation by Casagrande et~al.\ (2011, hereafter C11). They
derived improved stellar parameters and ages of all stars in the
Geneva-Copenhagen Survey (GCS, \citealt{nordstrom04}). We selected a
small sub-sample of 47 GCS stars from C11. These stars have ages with
errors smaller than 16\,\% if young ($<2.8$\,Gyr) and smaller than
20\,\% otherwise ($>2.8$\,Gyr), as given by C11. We used the effective
temperatures and metallicities given by C11 along with the stars'
visual magnitudes and Hipparcos parallaxes to derive ages using our
isochrone-fitting program, i.e., we used the exact same input data
employed by C11. Figure \ref{fig:dasilvagcs}b shows the comparison of
C11 and our ages for this small sample. The agreement between the two
sets of ages is excellent, with a mean difference of only
$0.03\pm0.45$\,Gyr. A closer inspection of this Figure suggests very
small systematic differences. For stars older than 7\,Gyr, our ages
are slightly younger ($-0.48\pm0.33$\,Gyr) whereas for stars younger
than 4\,Gyr they appear slightly old (although the mean difference is
still consistent with zero: $0.2\pm0.4$\,Gyr).  In this test we employed 
the ages from C11 derived using Padova isochrones. A similar test using
the ages they derived with BASTI isochrones (e.g.,
\citealt{basti04,basti06}) gives almost the same result. Thus, we show
again that statistical biases and the choice of isochrone grid appears
to have a minor impact on our project.

\section{Summary and Conclusions}

We have started a program that takes advantage of wide binaries with
evolved components and solar-type main-sequence companions in order to
obtain constraints for the calibration of age-rotation relations
(gyrochronology) at ages of several Gyr. We have mined published
catalogs of wide binaries and assembled a sample of 38 wide binary
systems with either turnoff/subgiant (20) or white dwarf (18)
components that are well positioned to provide ages for these
systems. For an initial subsample comprised of 15 of the binaries with
turnoff or subgiant components we measured precise stellar parameters
from high resolution optical spectra, and used these results to derive
ages for the binary systems using theoretical isochrones. As judged
from a one-to-one comparison with results obtained by state-of-the-art
analysis that takes into account a number of known systematics of the
isochrone technique, our own isochronal ages and the adopted
uncertainties appear to be robust. We obtained the ages of 16 of our
18 systems with white dwarf components directly from the literature,
having been derived from models of white dwarf cooling and
empirically-calibrated initial-to-final-mass relationships.

The ages of the 31 wide binary systems provided in this work range
from 1 to 9 Gyr. At least 15 of these systems have $1\sigma$
uncertainties of about 0.5 Gyr or less. Those for which the rotation
period of the main sequence companion is suitable to be measured will
thus provide useful constraints for gyrochronology relations, in a
regime of ages where the only constraint currently available is that
provided by the Sun. Observations aimed at the measurement of the
rotation periods of these stars have started and will be reported in a
future paper.

As larger and more varied samples of wide binaries start to become
available thanks to ongoing and planned all-sky surveys, the potential
of programs like the one presented in this paper for the purposes of
gyrochronology will only increase. Taking advantage of the SUPERBLINK
survey \citep{lepine08}, we have already more than doubled the sample
reported in Tables \ref{tab:hipp_doubles} to \ref{tab:WDs}, and
observations aimed to derive their ages are scheduled. With the help
of increasingly improving photometric distances and stellar
parameters, it may be possible today or in the near future to extend
this work to the fainter database of the Sloan Digital Sky Survey.
Finally, the unprecedented scale, depth, and precision levels of the
data expected from upcoming projects like {\it
  Gaia} and LSST will provide large quantities of wide binary systems
suitable not only for gyrochronology but for a large range of other
applications \citep{prague}.

\acknowledgments

We thank Jorge Mel\'endez for interpolating the Yale-Yonsei isochrones
and producing the very fine grid used in this work. We thank Sebastien
L\'epine for mining the SUPERBLINK survey for targets relevant for
gyrochronology, allowing a significant expansion of the present
program. Work by JC was supported by NASA through Hubble Fellowship
grant HST-HF-51239.01-A, awarded by the Space Telescope Science
Institute, which is operated by the Association of Universities for
Research in Astronomy, Inc., for NASA, under contract NAS5-26555.
I.R.'s work was performed under contract with the California Institute
of Technology (Caltech) funded by NASA through the Sagan Fellowship
Program. This work has made use of catalogs from the Astronomical Data
Center at NASA Goddard Space Flight Center, and the VizieR and SIMBAD
databases operated at CDS, Strasbourg, France.

\clearpage



\clearpage







\clearpage

\begin{deluxetable}{ccccccccccccccc}
\tabletypesize{\scriptsize}
\rotate
\tablewidth{0pc}
\tablecaption{Wide binaries with recently evolved primaries, both
  stars in {\it Hipparcos}.}
\tablehead{
\colhead{HIP}  
& \colhead{$\mu_{\rm RA}$}  
& \colhead{$\mu_{\rm DEC}$}  
& \colhead{$\pi$}  
& \colhead{$\sigma_{\pi}$}
& \colhead{$V$}
& \colhead{$B-V$}
& \colhead{$V-J$}
& \colhead{Sp.Type$^{1}$}
& \colhead{$\Delta\theta$}
& \colhead{$\Delta\mu$}
& \colhead{NLTT/LSPM}
& \colhead{Comments}
\\
\colhead{} & 
\colhead{(mas/yr)} & 
\colhead{(mas/yr)} & 
\colhead{(mas)} & 
\colhead{(mas)} & 
\colhead{(mag)} & 
\colhead{(mag)} & 
\colhead{(mag)} & 
\colhead{} &
\colhead{(arcsec)} & 
\colhead{(mas/yr)} & 
\colhead{} & 
\colhead{} &
& 
}
\startdata

  9243 &  137.5 &    125.9 &  7.81 & 0.79 &  8.27 & 0.88 & 1.64 & K0IV & 36.0 & 3.5 &  6664 & \\
  9247 &  138.0 &    129.4 &  7.82 & 0.94 &  9.11 & 0.54 & 1.02 & F8 & &  &  6665 & \\ \hline
 10305 &  378.2 &    -70.6 & 21.71 & 1.67 &  5.65 & 0.55 & 0.60 & F8V  & 16.7 & 5.5 &  7323 & \\
 10303 &  383.5 &    -72.2 & 20.89 & 7.39 &  7.56 & 0.72 & 1.03 & G4 & &  &  7321 & \\ \hline
 58240 & -174.0 &    -5.3 & 21.13 & 6.41 &  7.80 & 0.67 & 1.31 &G3V & 18.8 & 6.9 & 29039 & pre-MS star$^{1}$\\
 58241 & -180.9 &    -5.4 &  8.35 & 6.35 &  7.83 & 0.64 & 1.23 & G4V & &  & 29041 & \\ \hline
 74432 & -594.9 &  296.3 & 36.71 & 1.91 &  6.69 & 0.68 & 1.23 & G5V & 23.4 & 12.8 & 39601 & secondary within box\\
 74434 & -594.9 &  283.5 & 21.49 & 4.98 &  7.71 & 0.74 & 1.47 & G7V & &  & 39603 & \\ \hline
 81991 & -215.7 & -257.7 & 21.80 & 0.91 &  6.55 & 0.89 & 1.61 & G5 & 163.0 & 4.5 & 43483 & \\
 81988 & -219.3 & -255.0 & 23.06 & 1.73 & 10.31 & 1.04 & 2.01 & K3 & &  & 43481 & \\ \hline
 94076 &     51.7 &  186.1 & 19.11 & 1.57 &  6.70 & 0.64 & 1.15 & G1V & 15.9 & 20.3 & 47474 & spectroscopic binary$^{1}$\\
 94075 &     69.1 &  196.6 & 23.48 & 4.95 &  7.97 & 0.75 & 1.22 & G5 & &  & 47473 & substellar companion \\ \hline
101916 &  322.2 &    21.2 & 33.20 & 0.82 &  5.07 & 0.70 & 0.96 & G5IV & 214.0 & 4.5 & 49632 & \\
101932 &  317.8 &    20.1 & 34.90 & 1.14 &  8.52 & 0.91 & 1.58 & K2V & &  & 49639 & \\ \hline
101082 &  68  &  221  & 15.69 & 0.50 &  5.97 & 0.94 & 1.47 & K0III & 214.2 & 1.0 & J2029+8105 & \\
101166 &  68  &  220  & 14.51 & 0.70 &  8.68 & 0.64 & 1.13 & G5 & &  & J2030+8108 & \\ \hline
102532 & -30  & -198  & 25.82 & 1.20 &  4.25 & 1.04 & 0.97 & K1IV &  9.1 & 29.6 & J2046+1607E & \\
102531 & -1  & -204  & 26.35 & 2.40 &  4.97 & 0.49 & 2.91 & F7V & & &J2046+1607W & \\


\enddata

\tablecomments{(1) Obtained from the SIMBAD database.}

\label{tab:hipp_doubles}
\end{deluxetable}

\begin{deluxetable}{ccccccccccccccc}
\tabletypesize{\scriptsize}
\rotate
\tablewidth{0pc}
\tablecaption{Wide binaries with recently evolved primaries, only primary
  star in {\it Hipparcos}.}
\tablehead{
\colhead{HIP}  
& \colhead{NLTT/LSPM}
& \colhead{$\mu_{\rm RA}$}  
& \colhead{$\mu_{\rm DEC}$}  
& \colhead{$\pi$}  
& \colhead{$\sigma_{\pi}$}
& \colhead{$V$}
& \colhead{$B-V$}
& \colhead{$V-J$}
& \colhead{Sp.Type$^{1}$}
& \colhead{$\Delta\theta$}
& \colhead{$\Delta\mu$}
& \colhead{Comments}
\\
\colhead{} & 
\colhead{} &
\colhead{(mas/yr)} & 
\colhead{(mas/yr)} & 
\colhead{(mas)} & 
\colhead{(mas)} & 
\colhead{(mag)} & 
\colhead{(mag)} & 
\colhead{(mag)} & 
\colhead{} &
\colhead{(arcsec)} & 
\colhead{(mas/yr)} & 
\colhead{} &
& 
}
\startdata

  9148 &  6581       &   214  &    50  & 11.15 & 1.11 &   8.26 & 0.70 & 1.26 & G3V & 41.5 &    5   & \\
           &  6583       &   219  &    52  &           &         & 13.07 & 9.99 & 2.29 & &         &         & \\ \hline
 18824 & 12415       &   373  &   -12         &    19.35 & 0.63 &  6.80 & 0.63 & 1.17 & G1V      &    64.1 &    4    & \\
            & 12412       &   370  &   -12         &              & &   17.82 & 9.99 & 0.99 &       &      &      & white dwarf\\ \hline
 23926 & 14574       &  -73   &   289         &    18.69 & 0.49 &    6.75 & 0.66 & 1.24 & G3V  &    10.0 &   16    & \\
            & 14573       &  -64   &   276         & 18.46 & 0.65     &   10.11 & 0.67 & 1.62 &  G0 &         &         &HIP 23923 \\ \hline
 32935 & 17118       &   123  &   156         &     8.19  & 0.57 &    8.12 & 0.94 & 1.74 & K0IV &    46.1 &    2    & \\
            & 17117       &   123  &   154         &              &      &   10.11 & 0.55 & 1.09 & G0 &         &         & \\ \hline
114855 & 56282       &   370  &   -17        &    21.77 & 0.29 &    4.24 & 1.11 & 1.98 & K0III &    49.6 &   14    &giant; substellar companion \\
             & 56278       &   370  &   -17        &              &      &    9.73 & 1.06 & 2.42 & K3 &         &         & \\ \hline
115126 & 56466       &   266  &   -45        &    47.35 & 2.47 & 5.20 & 0.79 & 1.38 & G8.5IV &    12.2 &   32    & \\
             & 56465       &   266  &   -45        &              &        &    8.88 & 0.88 & 2.84 & K2V &         &         & \\ \hline
 18448 & J0356+6950  &  190 & -168 &     6.44 & 1.20 &    9.34 & 0.93 & 1.68 & K0 &    24.7 &   1 & \\
            & J0356+6951  &  189 & -167 &             &      &   14.72            & 9.99 & 2.52 &  &         &      & \\ \hline
  6914 & J0129+7412W &  194 & -121 &  17.89 & 0.60 &    7.26 & 0.67 & 1.23 & G5 &    27.0 &   42 & \\
           & J0129+7412E &  183 & -80 &             &      &   12.61 & 1.17 & 2.89 &  &         &         & \\ \hline
 20197 & J0419+1416  &  82 & -216 &   11.89 & 1.10 &    7.54 & 0.99 & 1.81 & G7III &   215.0 &   14 & \\
            & J0419+1419  &  87 & -203 &             &      &     13.12 & 0.75 & 2.74 &   &    &   & possibly M dwarf\\ \hline
107759 & J2149+1402S  &  159 & -41 &     5.78 & 1.50 &    9.65 & 0.74 & 1.41 & G5 &    13.3 &   9 & \\
             & J2149+1402N  &  154 & -33 &            &      &   99.99 & 9.99 & 9.99 &  &         &         & \\

\enddata

\tablecomments{(1) Obtained from the SIMBAD database.}

\label{tab:hipp_nonhipp}
\end{deluxetable}

\begin{deluxetable}{ccccccccccccccc}
\tabletypesize{\scriptsize}
\rotate
\tablewidth{0pc}
\tablecaption{Wide binaries with white dwarf components.}
\tablehead{
\colhead{NLTT}
& \colhead{$\mu_{\rm RA}$}  
& \colhead{$\mu_{\rm DEC}$}  
& \colhead{$\pi$}  
& \colhead{$\sigma_{\pi}$}
& \colhead{$V$}
& \colhead{$B-V$}
& \colhead{$V-J$}
& \colhead{Sp.Type$^{1}$}
& \colhead{$\Delta\theta$}
& \colhead{$\Delta\mu$}
& \colhead{Age}
& \colhead{Comments}
\\
\colhead{} & 
\colhead{(mas/yr)} & 
\colhead{(mas/yr)} & 
\colhead{(mas)} & 
\colhead{(mas)} & 
\colhead{(mag)} & 
\colhead{(mag)} & 
\colhead{(mag)} & 
\colhead{} &
\colhead{(arcsec)} & 
\colhead{(mas/yr)} & 
\colhead{(Gyr)} &
\colhead{} &
& 
}
\startdata

 1762 & 207  &  -53  &       &      &  16.59 & 0.30 & -0.078 & DA &  28.8 & 18 & 1.2$^{+0.15}_{-0.14}$ &           \\
 1759 & 222  &  -44  &  9.52 & 1.63 &  10.28 & 0.78 &  1.475 & K1 &       &    &                       & HIP 2600  \\ \hline
29967 & -299 & -328  &       &      &  17.26 & 9.99 &  1.19  & DA & 202.8 & 15 &                       &           \\
29948 & -292 & -313  & 22.94 & 1.63 &   9.96 & 0.99 &  1.89  & K4  &       &    &                       & HIP 59519 \\ \hline
13599 & 228  & -155  &       &      &  15.94 & 0.65 &  1.34  & DA/DC & 123.9 &  8 &   5.4$^{+2.6}_{-1.3}$ &           \\
13601 & 233  & -149  & 56.02 & 1.21 &   8.42 & 1.10 &  2.47  & K7 &       &    &                       & HIP 21482 \\ \hline
55288 & 548  &  -57  &       &      &  16.50 & 0.42 &  0.87  & DA &  41.8 &  9 &   3.6$^{+2.1}_{-0.9}$ &           \\
55287 & 551  &  -49  & 28.72 & 1.30 &   8.03 & 0.66 &  1.22  & G7 &       &    &                       & HIP 113231\\ \hline
44348 &  -92 &  252  &       &      &  17.50 & 0.10 &  9.99  & DA/DF &  28.7  &  8 &                       &           \\
44344 &  -88 &  259  & 18.18 & 1.35 &  11.46 & 1.02 &  2.34  & K7 &       &    &                       & phot plx  \\ \hline
 7890 & -116 & -145  &       &      &  17.39 &-1.19 &  1.24  & DA & 40.5 & 10 &   5.0$^{+2.5}_{-1.2}$ &           \\                   
 7887 & -119 & -155  & 22.22 & 2.50 &   9.84 & 1.01 &  1.90  & K3 &       &    &                       & phot plx  \\ \hline
 1374 &   72 & -205  &       &      &  16.22 & 0.36 &  0.17  & DA & 59.2 &  2 &   2.3$^{+1.6}_{-0.7}$ &           \\
 1370 &   74 & -206  &  8.00 & 0.65 &  12.90 & 1.70 &  2.16  & K6 &       &    &                       & phot plx  \\ \hline
LP378-537   &       &       &        &       & 16.2  &        &       & DA    &     &  & $1.0^{+0.2}_{-0.1}$ & \\
BD+23 2539  & -101  &  79   &  56.7  &       &  9.7  & 0.7422 & 1.252 & K0    &  20 &  &                     & \\ \hline
LP786-6     & -190  &  33   &        &       & 15.7  & -0.118 &-0.552 & DB    &     &  & $2.0^{+0.8}_{-0.6}$ & NLTT 20260 \\
BD-18 2482  &       &       &  29.49 &       & 12.8  & 1.0362 &       & K3    &  31 &  &                     & \\ \hline
40 Eri B    & -2228 & -3377 &        &       &  9.5  & 0.11   &-0.349 & DA    &  82 &45  & $5.0^{+1.1}_{-1.0}$ & NLTT 12868 \\  
40 Eri A    & -2240 & -3420 & 200.62 & 0.23  &  4.4  & 0.82   & 1.397 & K1    &     &  &                     & NLTT 12863 \\ \hline
L481-60     &  -415 &  -214 &  65.6  &       & 12.8  & 0.30   &       & DA    &     &1  & $1.1^{+0.2}_{-0.2}$ & NLTT 41169 \\
CD-37 10500 &  -415 &  -215 &  65.13 & 0.40  &  6.0  & 0.72   & 1.051 & G7IV  &  15 &  &                     & NLTT 41167 \\ \hline
L182-61     &   -46 &  -316 &        &       & 14.1  & -0.09  &-0.214 & DB    &     &1  & $1.4^{+1.0}_{-1.0}$ & NLTT 16355 \\
CD-59 1275  &   -46 &  -316 &  26.72 & 0.29  &  6.4  & 0.59   & 1.024 & G0    &  41 &  &                     & NLTT 16354 \\ \hline
CD-38 10980 &    76 &     1 &  76.00 & 2.56  & 11.0  & -0.14  &-0.548 & DA    &     &6  & $1.4^{+0.9}_{-0.5}$ & HIP 80300 \\
CD-38 10983 &    72 &     5 &  78.26 & 0.37  &  5.4  & 0.597  & 0.966 & G5    & 345 &  &                     & HIP 80337 \\ \hline
LHS300B     &       &       &        &       & 17.8  &        &       & DB    &     &  & $7.9^{+1.1}_{-2.4}$ & \\
LHS300A     &       &       &  32.3  &       & 13.2  &        &       & K     &     &  &                     & \\ \hline
LP592-80    &       &       &        &       & 17.2  &        &       & DA    &  49 &  & $2.6^{+1.3}_{-1.3}$ & \\ 
BD-1 469    &   253 &   -60 &  14.89 & 0.84  &  5.4  & 1.04   & 1.966 & K1IV  &     &  &                     & HIP 15383 \\ \hline
G216-B14B   &       &       &        &       & 15.5  &        &       & DA    &  23 &  & $1.2^{+0.1}_{-0.2}$ & B-band mag \\
G216-B14A   &       &       &        &       & 12.0  &        & 2.864 &       &     &  &                     & \\ \hline
G116-16     &    38 &  -296 &  34.6  & 4.00  & 15.4  & 0.21   & 0.355 & DA    &1020 &17  & $1.6^{+0.2}_{-0.2}$ & NLTT 21338 \\
BD+44 1847  &    34 &  -279 &  19.0  & 1.04  &  9.0  & 0.68   & 1.315 & G0    &     &  &                     & NLTT 21283 \\ \hline
G273-B1B    &   169 &     7 &        &       & 16.4  &        &       & DA    &  36 &  & $3.4^{+0.8}_{-0.7}$ & \\ 
G273-B1A    &       &       &        &       & 11.0  &        &          & G     &     &  &                     & B-band mag \\ \hline

\enddata

\tablecomments{(1) Spectral types are taken preferentially from the original
  source, i.e., either \citet{garces11} or \citet{zhao11}, or else from the
  SIMBAD database.}

\label{tab:WDs}
\end{deluxetable}

\begin{table}
\caption{Observing Log and Derived Radial Velocities}
\centering\footnotesize
\begin{tabular}{lccrcr}\hline\hline
\input{table4.tex}
\end{tabular}
\label{t:obslog}
\end{table}

\begin{deluxetable}{cccrc}
\tabletypesize{\footnotesize}
\tablewidth{10.5cm}
\tablecaption{Measured Stellar Parameters}
\tablehead{\colhead{Object} & \colhead{$\teff$ (K)} & \colhead{$\logg$ [cgs]} & \colhead{$\feh$} & Method\tablenotemark{1}}
\startdata
\input{table5.tex}
\enddata
\tablenotetext{1}{
(1) Excitation/ionization balance of iron lines, (2) Photometric
$\teff$ and parallax, (3) Photometric $\teff$ and ionization balance,
(4) \ha\ $\teff$ and parallax, (5) \ha\ $\teff$ and ionization
balance.} 
\label{t:pars}
\end{deluxetable}

\begin{deluxetable}{lccrc}
\tabletypesize{\footnotesize}
\tablewidth{10.5cm}
\tablecaption{Adopted Stellar Parameters}
\tablehead{\colhead{Object} & \colhead{$\teff$ (K)} & \colhead{$\logg$ 
[cgs]\tablenotemark{1}} & \colhead{$\feh$} & \colhead{$\logg_{\rm iso}$ 
[cgs]\tablenotemark{2}}}
\startdata
\input{table6.tex}
\enddata
\tablenotetext{1}{Average surface gravity obtained as described in 
Section \ref{adopted}.}
\tablenotetext{2}{Preferred value. Surface gravity obtained using 
$\teff$ and $\feh$ from this Table, observed visual magnitudes, 
Hipparcos parallaxes, and our isochrone technique described in Section \ref{ages}.}
\label{t:final_pars}
\end{deluxetable}

\begin{deluxetable}{ccccccccccccccc}
\tabletypesize{\scriptsize}
\footnotesize
\tablewidth{0pc}
\tablecaption{Isochronal ages for wide binaries with turnoff and subgiant
components}
\tablehead{
\colhead{ID}
& \colhead{this work}  
& \colhead{da Silva et al. (2006)}  
\\
\colhead{} & 
\colhead{(Gyr)} & 
\colhead{(Gyr)} & 
& 
}
\startdata

HIP 9148         &  6.75$^{+1.57}_{-1.32}$ & 7.776 $\pm$ 1.744 & \\
NLTT 6583        &                         &                   & \\ \hline
HIP 9243         &  3.25$^{+0.67}_{-0.59}$ & 2.790 $\pm$ 0.423 & \\
HIP 9247         &  5.25$^{+0.11}_{-1.20}$ & 5.594 $\pm$ 0.995 & \\ \hline
HIP 10305        &  2.75$^{+0.50}_{-0.33}$ & 2.587 $\pm$ 0.326 & \\
HIP 10303        &  7.75$^{+0.62}_{-2.03}$ & 4.832 $\pm$ 3.710 & \\ \hline
HIP 18824        &  4.75$^{+0.47}_{-0.21}$ & 4.672 $\pm$ 0.483 & \\
NLTT 12412       &                         &                   &  \\ \hline
HIP 20197        &  6.25$^{+2.55}_{-1.04}$ & 5.559 $\pm$ 1.573 & \\
LSPM J0419+1419  &                         &                   & \\ \hline
HIP 23926        &  4.75$^{+0.17}_{-0.57}$ & 4.674 $\pm$ 0.332 & \\
HIP 23923        &  5.75$^{+6.92}_{-2.34}$ & 4.381 $\pm$ 3.980 & \\ \hline
HIP 32935        &  8.75$^{+4.00}_{-2.07}$ & 7.891 $\pm$ 2.262 & \\
NLTT 17117       &  9.75$^{+1.29}_{-3.13}$ & 8.665 $\pm$ 3.655 & \\ \hline
HIP 74432        &  7.75$^{+1.39}_{-1.78}$ & 7.968 $\pm$ 2.045 & \\
HIP 74434        &  2.25$^{+4.92}_{-0.77}$ & 8.384 $\pm$ 3.618 & \\ \hline
HIP 81991        &  4.75$^{+3.56}_{-0.06}$ & 5.949 $\pm$ 1.636 & \\
HIP 81988        &                         &                   & \\ \hline
HIP 94076        &  4.25$^{+0.47}_{-0.66}$ & 3.490 $\pm$ 0.684 & \\
HIP 94075        &  8.75$^{+1.32}_{-2.87}$ & 4.463 $\pm$ 3.920 & \\ \hline
HIP 101916       &  3.25$^{+0.19}_{-0.22}$ & 2.944 $\pm$ 0.105 & \\
HIP 101932       &  6.25$^{+4.75}_{-3.13}$ & 3.655 $\pm$ 3.632 & \\ \hline
HIP 102532       &  1.25$^{+0.21}_{-0.32}$ & 1.114 $\pm$ 0.128 & \\
HIP 102531       &  2.75$^{+0.10}_{-0.77}$ & 1.774 $\pm$ 0.154 & \\ \hline
HIP 107759       &  6.25$^{+5.70}_{-1.72}$ & 7.130 $\pm$ 2.977 & \\
LSPM J2149+1402N &                         &                   & \\ \hline
HIP 115126       &  6.25$^{+0.93}_{-0.65}$ & 6.115 $\pm$ 0.869 & \\
NLTT 56465       &  3.25$^{+5.67}_{-1.32}$ & 3.699 $\pm$ 3.563 & \\ \hline
HIP 114855       &  1.25$^{+0.16}_{-0.34}$ & 1.460 $\pm$ 0.260 & \\
NLTT 56278       &                         & 5.021 $\pm$ 3.957 & \\

\enddata

\label{tab:ages}
\end{deluxetable}




\end{document}

%% file: table4.tex
Object    & UT date      & UT time    & Exp.\ time & S/N                 & Radial velocity    \\
          & (yyyy-mm-dd) & (hh:mm:ss) & (seconds)  & $\lambda=6000$\,\AA & (\kms) 	    \\ \hline
Hebe	  & 2010-09-21   & 04:10:27   & 1004	   & 289	         &   $\ldots$	      \\
HIP9148	  & 2010-09-22   & 03:38:02   & 3648	   & 420	         & $ 17.13 \pm 0.25$  \\
HIP9243	  & 2010-09-21   & 04:47:48   & 2670	   & 447	         & $  7.84 \pm 0.24$  \\
HIP9247	  & 2010-09-21   & 05:38:55   & 3250	   & 314	         & $  7.26 \pm 0.50$  \\
HIP10305  & 2011-01-04   & 01:43:37   &  138	   & 479	         & $ -3.81 \pm 0.78$  \\
HIP10303  & 2011-01-04   & 01:05:54   &  843	   & 505	         & $ -4.90 \pm 0.25$  \\
HIP18824  & 2010-09-21   & 08:48:35   & 2081	   & 111	         & $ 10.64 \pm 0.31$  \\
HIP20197  & 2010-09-22   & 09:17:40   &  405	   & 437	         & $-18.58 \pm 0.31$  \\
HIP23926  & 2010-09-21   & 06:48:54   &  435	   & 409	         & $ 44.30 \pm 0.30$  \\
HIP23923  & 2010-09-21   & 07:08:18   & 4606	   & 180	         & $ 45.27 \pm 0.22$  \\
HIP32935  & 2010-09-22   & 07:00:49   & 1725	   & 378	         & $ 88.24 \pm 0.32$  \\
NLTT17117 & 2010-09-22   & 08:42:38   & 1800	   & 236	         & $ 88.61 \pm 0.40$  \\
HIP74432  & 2010-09-21   & 23:16:51   &  180	   & 211	         & $-39.03 \pm 0.40$  \\
HIP74434  & 2010-09-21   & 23:23:45   &  120	   & 120	         & $-38.56 \pm 0.36$  \\
HIP81991  & 2010-09-20   & 23:31:06   &  150	   & 270	         & $ -5.29 \pm 0.25$  \\
HIP94076  & 2010-09-21   & 23:32:49   &  947	   & 439	         & $-44.08 \pm 0.43$  \\
HIP94075  & 2010-09-21   & 23:55:35   & 2716	   & 548	         & $-40.08 \pm 0.41$  \\
HIP101916 & 2010-09-21   & 00:54:04   &   90	   & 513	         & $-53.89 \pm 0.29$  \\
HIP101932 & 2010-09-21   & 01:02:19   & 1200	   & 451	         & $-52.43 \pm 0.24$  \\
HIP102532 & 2010-09-22   & 00:52:17   &   30	   & 442	         & $ -5.72 \pm 0.32$  \\
HIP102531 & 2010-09-22   & 00:59:06   &   75	   & 443	         & $ -6.56 \pm 0.49$  \\
HIP107759 & 2010-09-22   & 01:12:56   & 3112	   & 286	         & $-63.95 \pm 0.33$  \\
HIP115126 & 2010-09-21   & 03:02:41   &  190	   & 436	         & $  8.49 \pm 0.26$  \\
NLTT56465 & 2010-09-21   & 03:16:26   & 1478	   & 453	         & $ 11.55 \pm 0.19$  \\
HIP114855 & 2010-09-21   & 01:52:46   &  105	   & 624	         & $-25.62 \pm 0.28$  \\
NLTT56278 & 2010-09-21   & 01:58:50   & 2400	   & 276	         & $-22.88 \pm 0.36$  \\ \hline

%% file: table5.tex
hip9148             & $5649\pm 27$ & $4.01\pm0.03$ & $-0.146\pm0.020$ & (1) \\
                    & $5619\pm 25$ & $3.93\pm0.06$ & $-0.165\pm0.022$ & (2) \\
                    & $5619\pm 24$ & $3.94\pm0.05$ & $-0.164\pm0.022$ & (3) \\
                    & $5622\pm 31$ & $3.88\pm0.06$ & $-0.172\pm0.024$ & (4) \\
                    & $5620\pm 31$ & $3.94\pm0.05$ & $-0.164\pm0.022$ & (5) \\ \hline
hip9243             & $5045\pm 17$ & $3.41\pm0.07$ & $-0.076\pm0.025$ & (1) \\
                    & $5048\pm 45$ & $3.46\pm0.06$ & $-0.068\pm0.025$ & (4) \\
                    & $5048\pm 45$ & $3.43\pm0.06$ & $-0.077\pm0.025$ & (5) \\ \hline
hip9247             & $6135\pm 39$ & $4.12\pm0.06$ & $-0.193\pm0.025$ & (1) \\
                    & $6130\pm 24$ & $4.20\pm0.08$ & $-0.188\pm0.030$ & (2) \\
                    & $6130\pm 24$ & $4.10\pm0.06$ & $-0.193\pm0.025$ & (3) \\
                    & $6131\pm 41$ & $4.18\pm0.08$ & $-0.190\pm0.028$ & (4) \\
                    & $6140\pm 41$ & $4.13\pm0.06$ & $-0.190\pm0.025$ & (5) \\ \hline
hip10305            & $6131\pm 46$ & $4.13\pm0.06$ & $ 0.058\pm0.027$ & (1) \\
                    & $5996\pm 82$ & $3.80\pm0.05$ & $-0.022\pm0.041$ & (2) \\
                    & $5996\pm 82$ & $3.84\pm0.11$ & $-0.020\pm0.039$ & (3) \\
                    & $6137\pm 25$ & $3.86\pm0.04$ & $ 0.035\pm0.057$ & (4) \\
                    & $6110\pm 25$ & $4.09\pm0.06$ & $ 0.046\pm0.027$ & (5) \\ \hline
hip10303            & $5719\pm 14$ & $4.33\pm0.04$ & $ 0.112\pm0.015$ & (1) \\
                    & $5727\pm 23$ & $4.31\pm0.18$ & $ 0.111\pm0.016$ & (2) \\
                    & $5727\pm 23$ & $4.35\pm0.05$ & $ 0.114\pm0.015$ & (3) \\
                    & $5739\pm 24$ & $4.31\pm0.18$ & $ 0.110\pm0.023$ & (4) \\
                    & $5742\pm 24$ & $4.40\pm0.06$ & $ 0.125\pm0.017$ & (5) \\ \hline
hip18824            & $5880\pm 26$ & $4.07\pm0.05$ & $-0.066\pm0.027$ & (1) \\
                    & $5961\pm108$ & $3.93\pm0.06$ & $-0.042\pm0.082$ & (4) \\
                    & $5957\pm110$ & $4.25\pm0.07$ & $-0.025\pm0.030$ & (5) \\ \hline
hip20197            & $4845\pm 26$ & $3.24\pm0.06$ & $-0.035\pm0.028$ & (1) \\
                    & $4864\pm 49$ & $3.34\pm0.08$ & $-0.022\pm0.028$ & (4) \\
                    & $4865\pm 50$ & $3.30\pm0.07$ & $-0.031\pm0.027$ & (5) \\ \hline
hip23926            & $5743\pm 24$ & $3.85\pm0.07$ & $-0.149\pm0.026$ & (1) \\
                    & $5703\pm 61$ & $3.81\pm0.02$ & $-0.168\pm0.029$ & (2) \\
                    & $5702\pm 61$ & $3.74\pm0.07$ & $-0.174\pm0.029$ & (3) \\
                    & $5706\pm 32$ & $3.81\pm0.02$ & $-0.168\pm0.028$ & (4) \\
                    & $5711\pm 31$ & $3.77\pm0.07$ & $-0.170\pm0.028$ & (5) \\ \hline
hip23923            & $4963\pm110$ & $4.61\pm0.03$ & $-0.171\pm0.079$ & (4) \\
                    & $4966\pm110$ & $4.69\pm0.14$ & $-0.130\pm0.085$ & (5) \\ \hline
hip32935            & $4853\pm 16$ & $3.09\pm0.04$ & $-0.308\pm0.023$ & (1) \\
                    & $4916\pm 67$ & $3.23\pm0.08$ & $-0.281\pm0.024$ & (4) \\
                    & $4915\pm 67$ & $3.23\pm0.05$ & $-0.284\pm0.024$ & (5) \\ \hline
nltt17117           & $5835\pm 27$ & $4.24\pm0.07$ & $-0.405\pm0.030$ & (1) \\
                    & $5855\pm 51$ & $4.36\pm0.07$ & $-0.382\pm0.033$ & (2) \\
                    & $5856\pm 51$ & $4.29\pm0.08$ & $-0.390\pm0.031$ & (3) \\
                    & $5772\pm 57$ & $4.32\pm0.07$ & $-0.422\pm0.047$ & (4) \\
                    & $5802\pm 58$ & $4.17\pm0.07$ & $-0.424\pm0.030$ & (5) \\ \hline
hip74432            & $5686\pm 23$ & $4.13\pm0.05$ & $ 0.068\pm0.025$ & (1) \\
                    & $5690\pm 74$ & $4.28\pm0.05$ & $ 0.093\pm0.040$ & (2) \\
                    & $5686\pm 75$ & $4.12\pm0.05$ & $ 0.074\pm0.025$ & (3) \\
                    & $5742\pm 49$ & $4.32\pm0.04$ & $ 0.108\pm0.034$ & (4) \\
                    & $5750\pm 49$ & $4.26\pm0.07$ & $ 0.099\pm0.030$ & (5) \\ \hline
hip74434            & $5573\pm 35$ & $4.30\pm0.10$ & $ 0.087\pm0.040$ & (1) \\
                    & $5573\pm 66$ & $4.11\pm0.13$ & $ 0.064\pm0.056$ & (2) \\
                    & $5577\pm 64$ & $4.32\pm0.10$ & $ 0.084\pm0.040$ & (3) \\
                    & $5737\pm 86$ & $4.24\pm0.13$ & $ 0.106\pm0.090$ & (4) \\
                    & $5701\pm 84$ & $4.61\pm0.10$ & $ 0.137\pm0.053$ & (5) \\ \hline
hip81991            & $5030\pm 16$ & $3.53\pm0.04$ & $-0.243\pm0.026$ & (1) \\
                    & $5061\pm 64$ & $3.60\pm0.06$ & $-0.232\pm0.026$ & (2) \\
                    & $5061\pm 63$ & $3.60\pm0.04$ & $-0.235\pm0.026$ & (3) \\
                    & $5187\pm 66$ & $3.66\pm0.04$ & $-0.223\pm0.057$ & (4) \\
                    & $5170\pm 64$ & $3.87\pm0.09$ & $-0.191\pm0.037$ & (5) \\ \hline
hip94076            & $5953\pm 46$ & $4.33\pm0.09$ & $ 0.151\pm0.031$ & (1) \\
                    & $5992\pm 20$ & $3.99\pm0.08$ & $ 0.134\pm0.089$ & (4) \\
                    & $5951\pm 20$ & $4.33\pm0.09$ & $ 0.152\pm0.031$ & (5) \\ \hline
hip94075            & $5619\pm 23$ & $4.46\pm0.05$ & $ 0.266\pm0.025$ & (1) \\
                    & $5489\pm 45$ & $4.40\pm0.10$ & $ 0.276\pm0.054$ & (2) \\
                    & $5491\pm 45$ & $4.16\pm0.07$ & $ 0.214\pm0.040$ & (3) \\
                    & $5632\pm 18$ & $4.47\pm0.10$ & $ 0.272\pm0.025$ & (4) \\
                    & $5630\pm 18$ & $4.49\pm0.05$ & $ 0.275\pm0.025$ & (5) \\ \hline
hip101916           & $5775\pm 14$ & $3.88\pm0.05$ & $ 0.068\pm0.016$ & (1) \\
                    & $5616\pm 63$ & $3.68\pm0.02$ & $ 0.020\pm0.052$ & (2) \\
                    & $5611\pm 64$ & $3.51\pm0.09$ & $-0.001\pm0.041$ & (3) \\
                    & $5660\pm 24$ & $3.71\pm0.01$ & $ 0.028\pm0.037$ & (4) \\
                    & $5666\pm 24$ & $3.62\pm0.06$ & $ 0.021\pm0.031$ & (5) \\ \hline
hip101932           & $5030\pm 33$ & $4.41\pm0.06$ & $-0.007\pm0.023$ & (1) \\
                    & $5075\pm 19$ & $4.62\pm0.02$ & $ 0.039\pm0.043$ & (2) \\
                    & $5075\pm 20$ & $4.52\pm0.04$ & $ 0.006\pm0.027$ & (3) \\
                    & $5092\pm 33$ & $4.62\pm0.02$ & $ 0.035\pm0.039$ & (4) \\
                    & $5095\pm 33$ & $4.56\pm0.04$ & $ 0.012\pm0.029$ & (5) \\ \hline
hip102532           & $4918\pm 33$ & $3.26\pm0.12$ & $ 0.142\pm0.056$ & (1) \\
                    & $4866\pm 47$ & $2.93\pm0.06$ & $ 0.061\pm0.074$ & (4) \\
                    & $4864\pm 46$ & $3.11\pm0.09$ & $ 0.133\pm0.055$ & (5) \\ \hline
hip102531           & $6403\pm 60$ & $3.91\pm0.07$ & $ 0.154\pm0.023$ & (1) \\
                    & $6277\pm 67$ & $3.77\pm0.03$ & $ 0.088\pm0.039$ & (2) \\
                    & $6276\pm 67$ & $3.65\pm0.08$ & $ 0.081\pm0.036$ & (3) \\
                    & $6297\pm 24$ & $3.77\pm0.03$ & $ 0.095\pm0.035$ & (4) \\
                    & $6301\pm 24$ & $3.69\pm0.08$ & $ 0.093\pm0.032$ & (5) \\ \hline
hip107759           & $5530\pm 15$ & $4.03\pm0.03$ & $-0.368\pm0.019$ & (1) \\
                    & $5437\pm 60$ & $3.86\pm0.12$ & $-0.417\pm0.031$ & (2) \\
                    & $5437\pm 60$ & $3.79\pm0.08$ & $-0.427\pm0.031$ & (3) \\
                    & $5512\pm 52$ & $3.84\pm0.12$ & $-0.403\pm0.033$ & (4) \\
                    & $5495\pm 53$ & $3.97\pm0.05$ & $-0.390\pm0.021$ & (5) \\ \hline
hip115126           & $5557\pm 22$ & $4.13\pm0.05$ & $ 0.185\pm0.027$ & (1) \\
                    & $5383\pm 60$ & $3.86\pm0.03$ & $ 0.142\pm0.047$ & (2) \\
                    & $5378\pm 60$ & $3.71\pm0.09$ & $ 0.110\pm0.046$ & (3) \\
                    & $5495\pm 30$ & $3.90\pm0.03$ & $ 0.135\pm0.038$ & (4) \\
                    & $5486\pm 29$ & $4.01\pm0.05$ & $ 0.153\pm0.028$ & (5) \\ \hline
nltt56465           & $5113\pm 56$ & $4.36\pm0.09$ & $ 0.181\pm0.036$ & (1) \\
                    & $5163\pm 27$ & $4.56\pm0.03$ & $ 0.213\pm0.046$ & (2) \\
                    & $5163\pm 27$ & $4.50\pm0.07$ & $ 0.192\pm0.039$ & (3) \\
                    & $5180\pm 32$ & $4.57\pm0.03$ & $ 0.208\pm0.044$ & (4) \\
                    & $5178\pm 32$ & $4.53\pm0.06$ & $ 0.196\pm0.040$ & (5) \\ \hline
hip114855           & $4762\pm 38$ & $2.76\pm0.04$ & $ 0.109\pm0.060$ & (4) \\
                    & $4766\pm 38$ & $2.79\pm0.15$ & $ 0.114\pm0.060$ & (5) \\ \hline
nltt56278           & $4769\pm 90$ & $4.58\pm0.02$ & $ 0.004\pm0.092$ & (4) \\
                    & $4772\pm 92$ & $4.42\pm0.16$ & $-0.040\pm0.082$ & (5) \\ \hline

%% file: table6.tex
hip9148   & $5625\pm35$ & $3.96\pm0.07$ & $-0.16\pm0.04$ & $3.90\pm0.08$ \\
hip9243   & $5046\pm35$ & $3.44\pm0.06$ & $-0.07\pm0.04$ & $3.49\pm0.08$ \\
hip9247   & $6132\pm35$ & $4.14\pm0.07$ & $-0.19\pm0.04$ & $4.15\pm0.11$ \\
hip10305  & $6109\pm69$ & $3.93\pm0.17$ & $ 0.03\pm0.05$ & $3.77\pm0.13$ \\
hip10303  & $5730\pm35$ & $4.35\pm0.06$ & $ 0.11\pm0.04$ & $4.30\pm0.08$ \\
hip18824  & $5893\pm68$ & $4.06\pm0.18$ & $-0.05\pm0.04$ & $3.90\pm0.07$ \\
hip20197  & $4854\pm36$ & $3.29\pm0.09$ & $-0.03\pm0.04$ & $3.32\pm0.09$ \\
hip23926  & $5716\pm36$ & $3.81\pm0.05$ & $-0.17\pm0.04$ & $3.88\pm0.08$ \\
hip23923  & $4964\pm79$ & $4.61\pm0.05$ & $-0.15\pm0.09$ & $4.62\pm0.13$ \\
hip32935  & $4875\pm64$ & $3.16\pm0.11$ & $-0.29\pm0.04$ & $3.21\pm0.14$ \\
nltt17117 & $5828\pm53$ & $4.28\pm0.11$ & $-0.40\pm0.04$ & $4.38\pm0.13$ \\
hip74432  & $5711\pm55$ & $4.23\pm0.12$ & $ 0.09\pm0.04$ & $4.30\pm0.07$ \\
hip74434  & $5599\pm87$ & $4.34\pm0.23$ & $ 0.09\pm0.05$ & $4.50\pm0.07$ \\
hip81991  & $5077\pm92$ & $3.61\pm0.11$ & $-0.23\pm0.04$ & $3.57\pm0.14$ \\
hip94076  & $5967\pm45$ & $4.19\pm0.26$ & $ 0.15\pm0.04$ & $3.98\pm0.13$ \\
hip94075  & $5587\pm86$ & $4.42\pm0.16$ & $ 0.26\pm0.04$ & $4.29\pm0.08$ \\
hip101916 & $5685\pm83$ & $3.70\pm0.11$ & $ 0.03\pm0.04$ & $3.70\pm0.07$ \\
hip101932 & $5073\pm41$ & $4.58\pm0.09$ & $ 0.01\pm0.04$ & $4.51\pm0.08$ \\
hip102532 & $4889\pm56$ & $3.02\pm0.19$ & $ 0.12\pm0.07$ & $2.98\pm0.13$ \\
hip102531 & $6307\pm61$ & $3.77\pm0.08$ & $ 0.11\pm0.05$ & $3.71\pm0.08$ \\
hip107759 & $5496\pm63$ & $3.98\pm0.11$ & $-0.40\pm0.04$ & $3.84\pm0.15$ \\
hip115126 & $5488\pm83$ & $3.93\pm0.13$ & $ 0.15\pm0.04$ & $3.90\pm0.07$ \\
nltt56465 & $5165\pm37$ & $4.54\pm0.07$ & $ 0.20\pm0.04$ & $4.50\pm0.07$ \\
hip114855 & $4764\pm35$ & $2.76\pm0.05$ & $ 0.11\pm0.05$ & $2.71\pm0.08$ \\
nltt56278 & $4770\pm66$ & $4.57\pm0.07$ & $-0.02\pm0.09$ & $4.52\pm0.08$ \\ \hline